\documentclass{IEEEtaes}

\usepackage{color,array,amsthm}
\usepackage{graphicx}
\usepackage{color}
\usepackage{lineno}
\usepackage{cite}
\usepackage{verbatim}
\usepackage{amsmath,amssymb,amsfonts}
\usepackage{algorithmic}
\usepackage{graphicx}
\usepackage{textcomp}
\usepackage{xcolor}
\usepackage{subfigure}
\usepackage{setspace}
\usepackage{svg}
\usepackage[ruled]{algorithm2e}
\usepackage{booktabs}
\usepackage{url}
\usepackage{diagbox}
\usepackage{xcolor}
\usepackage{multicol}
\usepackage{multirow}
\usepackage{tikz}% Make Orcid icon
\usepackage[linkcolor=blue,citecolor=blue,urlcolor=blue]{hyperref}
\definecolor{lime}{HTML}{A6CE39}
\DeclareRobustCommand{\orcidicon}{%
    \begin{tikzpicture}
    \draw[lime, fill=lime] (0,0) 
    circle [radius=0.16] 
    node[white] {{\fontfamily{qag}\selectfont \tiny ID}};    \draw[white, fill=white] (-0.0625,0.095) 
    circle [radius=0.007];    \end{tikzpicture}
    \hspace{-2mm}}
\foreach \x in {A, ..., Z}{%
    \expandafter\xdef\csname orcid\x\endcsname{\noexpand\href{https://orcid.org/\csname orcidauthor\x\endcsname}{\noexpand\orcidicon}}
    }

\jvol{XX}
\jnum{XX}
\jmonth{XXXXX}
\paper{1234567}
\pubyear{2024}
\doiinfo{TAES.2024.Doi Number}

\SetCommentSty{mycommfont}

\SetKwInput{KwInput}{Input}                % Set the Input
\SetKwInput{KwOutput}{Output}              % set the Output

\begin{document}

% Generation for 2.0x line width.
% \begin{spacing}{2.0}

%%%%%%%%%%%%%%%%%%%%%%%%%%%%%%%%%%%%%%%%%%%%%%%%%%%%%%%%%%%%%%%%%%%%%%%%%%%%%%%%%%
% Title
%%%%%%%%%%%%%%%%%%%%%%%%%%%%%%%%%%%%%%%%%%%%%%%%%%%%%%%%%%%%%%%%%%%%%%%%%%%%%%%%%%
\title{Generalizable Indoor Human Activity Recognition Method Based on Micro-Doppler Corner Point Cloud and Dynamic Graph Learning} 

%%%%%%%%%%%%%%%%%%%%%%%%%%%%%%%%%%%%%%%%%%%%%%%%%%%%%%%%%%%%%%%%%%%%%%%%%%%%%%%%%%
% Paper Infomation
%%%%%%%%%%%%%%%%%%%%%%%%%%%%%%%%%%%%%%%%%%%%%%%%%%%%%%%%%%%%%%%%%%%%%%%%%%%%%%%%%%
\author{XIAOPENG YANG\orcidA{}}
\affil{School of Information and Electronics, Beijing Institute of Technology, Beijing 100081, China\\Jiaxing Research Center of Beijing Institute of Technology, Jiaxing 314000, China} 

\author{WEICHENG GAO\orcidB{}}
\author{XIAODONG QU\orcidC{}}
\author{HAOYU MENG\orcidD{}}
\affil{School of Information and Electronics, Beijing Institute of Technology, Beijing 100081, China\\Key Laboratory of Electronic and Information Technology in Satellite Navigation, Beijing Institute of Technology, Beijing 100081, China} 

\receiveddate{Manuscript received January 8th, 2024; revised XXXXXXXX XXth, 2024; accepted XXXXXXXX XXth, 2024. Date of publication XXXXXXXX XXth, 2024; date of current version XXXXXXXX XXth, 2024.\\ \\
DOI. No. 10.1109/TAES.2024.XXXXXXX\\ \\
Refereeing of this contribution was handled by XXX XXX.\\ \\
This work was supported in part by the National Natural Science Foundation of China under Grant 62101042. This work was also supported in part by Beijing Institute of Technology Research Fund Program for Young Scholars under Grant XSQD-202205005. (Corresponding author: Xiaodong Qu.)}
%% \accepteddate{XXXXX XX XXXX}
%% \publisheddate{XXXXX XX XXXX}

\authoraddress{Author's addresses: Xiaopeng Yang, is with the School of Information and Electronics, Beijing Institute of Technology, Beijing 100081, China, and also with the Jiaxing Research Center of Beijing Institute of Technology, Jiaxing 314000, China, E-mail: (xiaopengyang@bit.edu.cn); Weicheng Gao, Xiaodong Qu, and Haoyu Meng, are with the School of Information and Electronics, Beijing Institute of Technology, Beijing 100081, China, and with the Key Laboratory of Electronic and Information Technology in Satellite Navigation, Beijing Institute of Technology, Beijing 100081, China, E-mail: (JoeyBG@126.com; xdqu@bit.edu.cn; menghaoyu0@163.com). \emph{(Corresponding Author: Xiaodong Qu.)}}

\markboth{AUTHOR ET AL.}{SHORT ARTICLE TITLE}
\maketitle

% Line numbers generation.
% \linenumbers
% \pagewiselinenumbers
% \switchlinenumbers

%%%%%%%%%%%%%%%%%%%%%%%%%%%%%%%%%%%%%%%%%%%%%%%%%%%%%%%%%%%%%%%%%%%%%%%%%%%%%%%%%%
% Abstract
%%%%%%%%%%%%%%%%%%%%%%%%%%%%%%%%%%%%%%%%%%%%%%%%%%%%%%%%%%%%%%%%%%%%%%%%%%%%%%%%%%
\begin{abstract}
Through-the-wall radar (TWR) human activity recognition can be achieved by fusing micro-Doppler signature extraction and intelligent decision-making algorithms. However, limited by the insufficient priori of tester in practical indoor scenarios, the trained models on one tester are commonly difficult to inference well on other testers, which causes poor generalization ability. To solve this problem, this paper proposes a generalizable indoor human activity recognition method based on micro-Doppler corner point cloud and dynamic graph learning. In the proposed method, DoG-$\mathbf{\mu}$D-CornerDet is used for micro-Doppler corner extraction on two types of radar profiles. Then, a micro-Doppler corner filtering method based on polynomial fitting smoothing is proposed to maximize the feature distance under the constraints of the kinematic model. The extracted corners from the two types of radar profiles are concatenated together into three-dimensional point cloud. Finally, the paper proposes a dynamic graph neural network (DGNN)-based recognition method for data-to-activity label mapping. Visualization, comparison and ablation experiments are carried out to verify the effectiveness of the proposed method. The results prove that the proposed method has strong generalization ability on radar data collected from different testers.\par
\end{abstract}


\IEEEpeerreviewmaketitle

%%%%%%%%%%%%%%%%%%%%%%%%%%%%%%%%%%%%%%%%%%%%%%%%%%%%%%%%%%%%%%%%%%%%%%%%%%%%%%%%%%
% Introduction
%%%%%%%%%%%%%%%%%%%%%%%%%%%%%%%%%%%%%%%%%%%%%%%%%%%%%%%%%%%%%%%%%%%%%%%%%%%%%%%%%%
% \newpage
\section{INTRODUCTION}
In recent years, the development of non-intrusive surveillance and monitoring systems has become crucial for applications in security, search and rescue operations, and situational awareness \cite{Main1,TMTT1,TMTT2,TMTT3}. Through-the-wall radar (TWR) technology emerges as a promising solution for detecting and recognizing human activities behind obstacles \cite{Main2}. However, the ability to accurately recognize and classify specific human activities through walls remains challenging tasks \cite{Songyongping, Main3}. Micro-Doppler refers to the modulation of radar signals by limbs and body parts \cite{Main4}. By analyzing the micro-Doppler signature, it is possible to extract unique motion features to identify and classify human activities.\par
\begin{figure*}
    \centering
    \includegraphics[width=\textwidth]{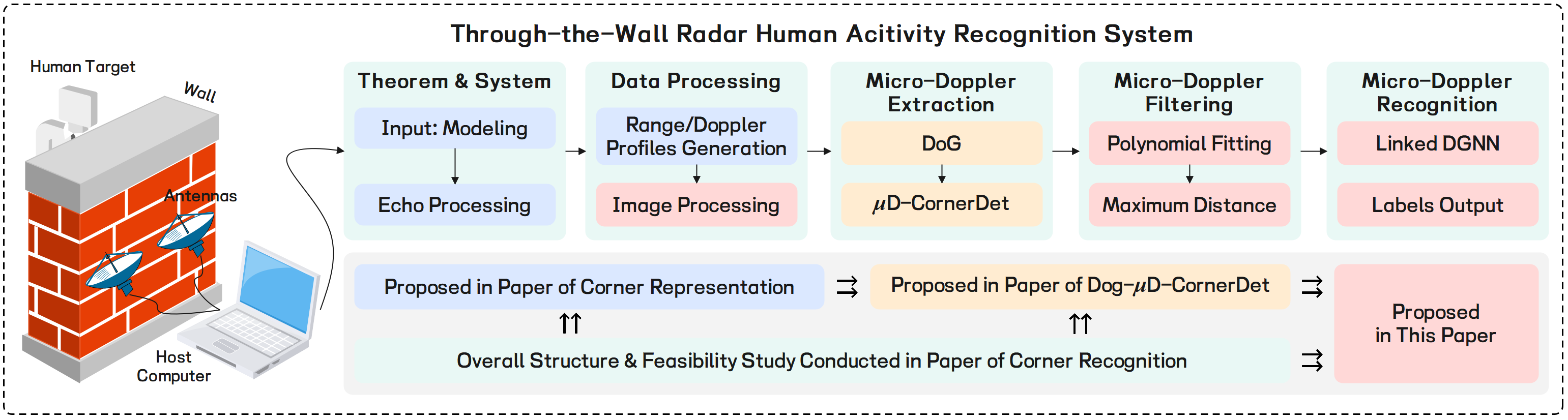}
    \caption{The full process of the proposed TWR human activity recognition system and the interrelationships of our previous works.}
    \label{Previous Works}
    \vspace{-0.0cm}
\end{figure*}
The processing flow of the human activity recognition system based on micro-Doppler analysis for TWR contains three common stages: signal and data processing, feature extraction and dimension reduction, and recognition decision \cite{System Stages}. A great deal of academic researches have been invested in all steps and a range of results have been achieved. Ram et. al., presented a simulation methodology for generating micro-Doppler radar signature of humans moving behind a wall \cite{Ram}. Liang, proposed a standard deviation (STD)-based approach to sense-through-wall and sense-through-wooden-door human detection, and made analysis on detection threshold selection \cite{Liang}. To cope with the issue that traditional time-frequency analysis algorithms were not good at displaying complete micro-Doppler information in radar echoes, a multiple Hilbert-Huang transform (MHHT) method was proposed for high-resolution time-frequency transform of digging fine-grained human activity hidden micro-Doppler signature in ultra-wideband (UWB) radar echoes during the through-wall detection environment \cite{Qi}. Sun et. al., demonstrated the effectiveness of a passive indoor human sensing technique using WiFi signals \cite{Sun}. These researches mainly focused on the first two steps: signal and data processing, feature extraction and dimension reduction.\par
With the rapid development of neural network algorithms in the field of computer vision, researchers began to focus on their applicability for radar image recognition tasks. Wang et. al., studied through-the-wall human activity classification using complex-valued CNN and graph conducted CNN \cite{Wang 1, Wang 2, Wang 3}. The methods were utilized to classify the human activity behind the wall with the input of concatenated time domain echo matrix. An et. al., proposed a method that cascaded robust principal component analysis (RPCA) and ResNet for through-the-wall radar human activity recognition, which was one of the pioneering works in the field combining feature separation and CNN \cite{An}. Other influential works were the series of Chen's cross-domain or cross-view learning methods \cite{Chen 1, Chen 2, Chen 3}. These networks achieved strong micro-Doppler distribution fitting capabilities by pre-collecting a large amount of data. However, the inference ability was heavily depended on the training data set. We have proposed a variety of methods for data augmentation \cite{TWR-MCAE}, finer-grained micro-Doppler signature extraction \cite{TWR-WSN-CRF} or faster inference speeds \cite{TWR-FMSN}, respectively. Unfortunately, the exsisted methods mentioned above were all trained, validated, and tested under the same tester. However, limited by the insufficient priori of tester in practical indoor scenarios, the trained models on one tester were commonly difficult to inference well on other testers, which caused poor generalization.\par
To address this issue, we have been dedicated in some researches works, shown in Fig. \ref{Previous Works}. The concept of micro-Doppler corner features was proposed and some theoretical analyses with front-end algorithms were given. In \cite{Journal of Radar, Micro-Doppler Corner Representation}, a feasibility validation for the task of recognition generalizations was presented, including kinematic models under walking and abnormal activities, and decision scheme. Detailed theoretical analysis of the micro-Doppler corner feature was developed, showing that the minimum number of corners needed to characterize human activity was $30$. DoG-$\mu$D-CornerDet was employed for corner extraction with strong robustness \cite{Micro-Doppler Corner Detection}. All these methodological and theoretical analyses would serve as a prelude to the full-process design ideas of the decision model and generalization recognition system proposed in this paper.\par
\begin{figure}
    \centering
    \includegraphics[width=0.48\textwidth]{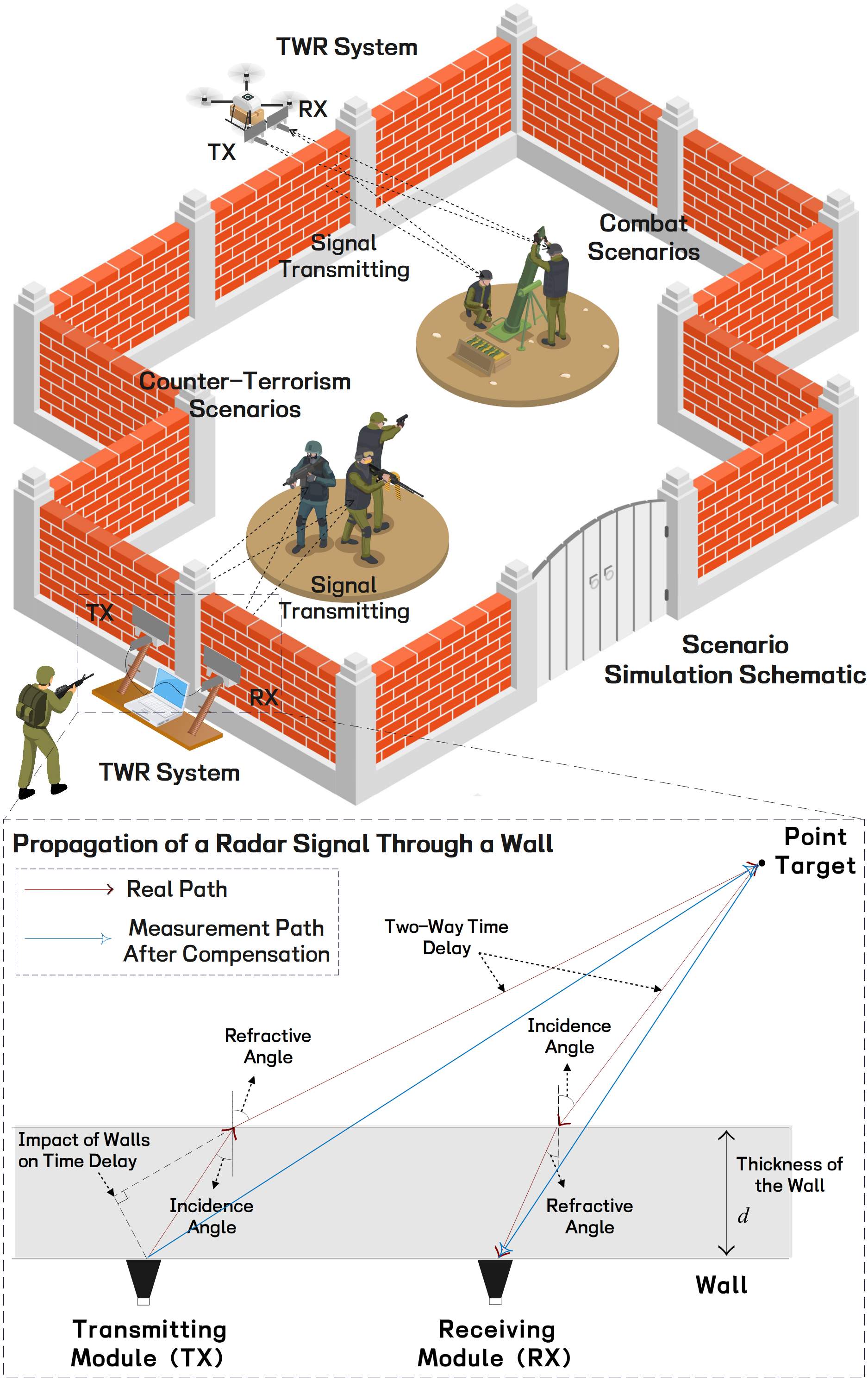}
    \caption{Schematic diagram of through-the-wall detection scene and electromagnetic signal propagation \cite{TWR System Diagram}.}
    \label{TWR System}
    \vspace{-0.0cm}
\end{figure}\par
Based on the previous works, there are two important issues that need to be addressed: First, how to filter the resulting corner features when the number of points detected by DoG-$\mu$D-CornerDet exceeds the constraints of the kinematic model. Second, how to design the recognition decision model applicable to the sparse micro-Doppler corner feature set. In response, this paper proposes a generalizable indoor human activity recognition method based on micro-Doppler corner point cloud and dynamic graph learning. Specifically, the contributions of this paper are as follows:\par
\textbf{(1) Micro-Doppler Corner Filtering Method:} A micro-Doppler corner point filtering method based on polynomial fitting smoothing is proposed to maximize the feature distance between classes. The corner points satisfy the constraints of the kinematic model. In addition, the corner point features on range and Doppler profiles are fused together to build a three-dimensional (3D) point cloud.\par
\textbf{(2) Micro-Doppler Corner Recognition Method:} A micro-Doppler corner point feature recognition method based on dynamic graph neural network (DGNN) is proposed, which can effectively achieve decision making on sparse 3D corner point cloud data, giving labeled results for indoor human activity.\par
% Fill in the blank with the structure of the main text, but in a brief way.
The rest of the paper is organized as follows. Section II gives the analysis of signal model and data preprocessing method. Section III gives the detailed design of the proposed method, including micro-Doppler extraction, filtering, and dynamic graph learning models. Section IV discusses the experiments and results. Section V gives the conclusion.\par

%%%%%%%%%%%%%%%%%%%%%%%%%%%%%%%%%%%%%%%%%%%%%%%%%%%%%%%%%%%%%%%%%%%%%%%%%%%%%%%%%%
% Main Text
%%%%%%%%%%%%%%%%%%%%%%%%%%%%%%%%%%%%%%%%%%%%%%%%%%%%%%%%%%%%%%%%%%%%%%%%%%%%%%%%%%
\section{SIGNAL MODEL AND DATA PREPROCESSING}
\subsection{Transceiver Module}
As shown in Fig. \ref{TWR System}, the human can be seen as extended target in the through-the-wall detection scenario. Without considering the multi-path effect, the radar echo can be approximated as the summation of the echoes from wall, human limb nodes, and background noise.\par
Assuming that the transmit signal contains $M$ pulse repetition intervals (PRI) during the coherent processing interval, then the transmitting signal in $m^{\mathrm{th}}$ PRI is:

\vspace{-0.3cm}
\begin{equation}
\begin{gathered}
S_{\mathrm{tx},m}(t)=A_{\mathrm{tx}}e^{j(2\pi(f_c(t-mT_s)+\frac{1}{2}\mu(t-mT_s)^2)+\varphi_{\mathrm{tx}})}\\mT_{s}\leq t\leq(m+1)T_{s}
\end{gathered},
\label{Transmitting Signal}
\end{equation}
where $A_\mathrm{tx}$ is the amplitude of the transmitted signal. $t_s$ is the PRI, $\mu=B/T_s$ is the slope of frequency modulation, $B$ is the bandwidth, and $f_c$ is the carrier frequency. $\varphi_\mathrm{tx}$ is the initial phase of the transmitted signal. $t$ in Eq. (\ref{Transmitting Signal}) denotes the fast time axis, $m$ in Eq. (\ref{Transmitting Signal}) denotes the slow time axis.\par
Given that the range resolution and wavelength for TWR is about at decimeter level \cite{Main2}, the number of human limb nodes is about $6$ to clearly represent the echo. For example, head, torso, left hand with arm, right hand with arm, left foot with leg, and right foot with leg are taken into consideration. Thus, the complete time-domain echo of the human target is:

\vspace{-0.3cm}
\begin{equation}
\begin{gathered}
S_{b,m}^{\prime}(t)=\sum_{i=1}^6S_{b,N_i,m}^{\prime}(t)\\=\sum_{i=1}^6A_{b,N_i}e^{j2\pi\left(f_c\tau_{N_i}^{\prime}(t)-\frac12\mu\tau_{N_i}^{\prime}(t)+\mu(t-mT_s)\tau_{N_i}^{\prime}(t)\right)}
\end{gathered}.
\end{equation}\par
Assuming that the wall echo is $S_{b,m,\mathrm{wall}}(t)$, and the background noise is $S_{b,m,\mathrm{noise}}(t)$, the complete radar time-domain echo can be eventually written as:

\vspace{-0.3cm}
\begin{equation}
\begin{gathered}
S_{b,m}(t)=\sum_{i=1}^6A_{b,N_i}e^{j2\pi\left(f_c\tau_{N_i}^{\prime}(t)-\frac12\mu\tau_{N_i}^{\prime}(t)+\mu(t-mT_s)\tau_{N_i}^{\prime}(t)\right)}\\+S_{b,m,\mathrm{wall}}(t)+S_{b,m,\mathrm{noise}}(t)
\end{gathered}.
\end{equation}\par
The first summation term of $S_{b,m}(t)$ contains the micro-Doppler information of the human body motion, including the distance and velocity of the nodes. Unfortunately, it is difficult to extract the distance and velocity information of the human motion directly from the radar time-domain echoes. Therefore, a combination of data processing algorithms such as time-frequency analysis and image transformation is needed to visualize the micro-Doppler signature.\par
\subsection{Data Preprocessing}
\begin{figure*}
    \centering
    \includegraphics[width=\textwidth]{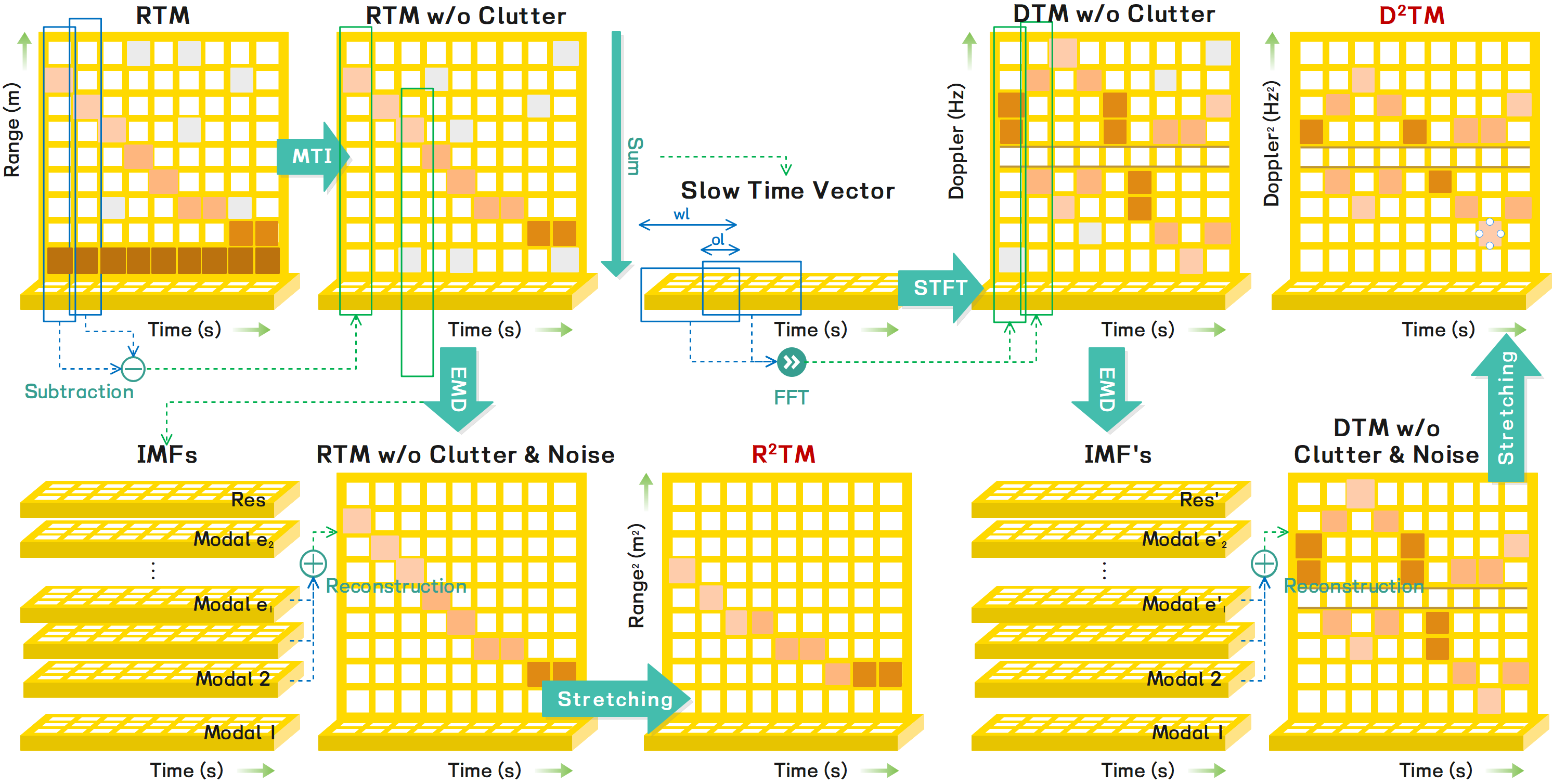}
    \caption{Schematic diagram of the data processing flow, where the two images with red titles correspond to the required range and Doppler profiles.}
    \label{Data Processing}
    \vspace{-0.0cm}
\end{figure*}

As shown in Fig. \ref{Data Processing}, we concatenate the radar time-domain echoes of all $M$ PRIs along the slow time dimension, take the modulus and normalize it to obtain the range-time map (RTM):

\vspace{-0.2cm}
\begin{equation}
\mathbf{RTM}=\operatorname{Norm}\left(\left|\operatorname{Con}_{m=0}^{M-1}\left(S_{b,m}(t)\right)\right|\right),
\end{equation}
where $\operatorname{Con}$ is a vector concatenation operation, and $\operatorname{Norm}$ is a linear normalization method for matrices. If the real-valued matrix to be normalized is $\mathbf{X}$, then:

\vspace{-0.2cm}
\begin{equation}
\mathrm{Norm}(\mathbf{X})=\frac{\mathbf{X}-\max(\mathbf{X})}{\max(\mathbf{X})-\min(\mathbf{X})},
\end{equation}
where $\max()$ and $\min()$ take the maximum and minimum values of the matrix, respectively. The RTM mainly contains the wall echo $\mathbf{RTM}_\mathrm{wall}$, the human motion echo $\mathbf{RTM}_\mathrm{mv}$, and the background noise $\mathbf{Ns}$, that is:

\vspace{-0.2cm}
\begin{equation}
\mathbf{RTM}=\mathbf{RTM}_\mathrm{wall}+\mathbf{RTM}_\mathrm{mv}+\mathbf{Ns}.
\label{RTM Equation}
\end{equation}\par
In this case, the term $\mathbf{RTM}_\mathrm{wall}$ can be removed by the moving target indicator (MTI) filter \cite{MTI}:

\vspace{-0.2cm}
\begin{equation}
\begin{gathered}
\mathbf{RTM_{mv}}+\mathbf{Ns}\approx\mathbf{RTM}[:,m+1]-\mathbf{RTM}[:,m]\\m=0,1,\cdots,M-1
\end{gathered}.
\label{Obtained RTM}
\end{equation}\par
\begin{figure*}[ht]
\begin{equation}
\mathbf{DTM_{mv}}+\mathbf{Ns}=\mathrm{Norm}\left|\mathrm{STFT}_{\mathrm{wl,ol}}\left(\sum_{t}\left(\mathrm{Con}_{m=0}^{M-1}\left(\mathrm{S}_{b,m}(t)\right)\left[:,m+1\right]-\mathrm{Con}_{m=0}^{M-1}\left(\mathrm{S}_{b,m}(t)\right)\left[:,m\right]\right)\right)\right|.
\label{DTM Equation}
\end{equation}
\hrulefill
\end{figure*}\par
\par
The Doppler information in the radar echoes is extracted by using short-time Fourier transform (STFT), generating Doppler-time map (DTM) \cite{STFT}. First, the time-domain echoes are concatenated along the slow time dimension. After suppressing the static target clutter by MTI, the matrix is summed along the fast time dimension to obtain a row vector of length $M$. The STFT is performed on this row vector and DTM is obtained by taking the amplitude normalization result of the transformed two-dimensional (2D) time-frequency matrix. The mathematical representation of the above process can be written in Eq. (\ref{DTM Equation}), where $\mathrm{wl},~\mathrm{ol}$ represent the window length of the STFT and the overlap length between neighboring windows, respectively.\par
The RTM in Eq. (\ref{Obtained RTM}) and DTM in Eq. (\ref{DTM Equation}) are both contaminated by background noise. Thus, the empirical modal decomposition (EMD) algorithm is used to further suppress the stationary target clutter and background noise that are not completely removed by MTI \cite{EMD}.\par
\begin{figure*}
    \centering
    \includegraphics[width=\textwidth]{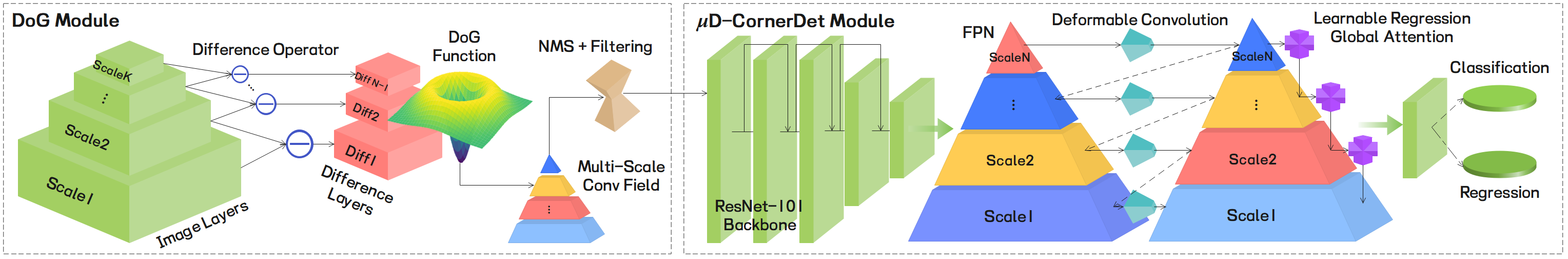}
    \caption{Simplified flowchart of the DoG-$\mu$D-CornerDet method.}
    \label{DoG-mD-CornerDet}
    \vspace{-0.0cm}
\end{figure*}\par
Finally, the vertical axes of $\mathbf{RTM}_\mathrm{mv}$ and $\mathbf{DTM}_\mathrm{mv}$ are stretched from linear to square units by interpolation. As a result, $\mathbf{R^2TM}$ and $\mathbf{D^2TM}$ are obtained, which are used as the foundations of micro-Doppler corner feature representation, and contain the squared distance and squared velocity information of the human limb nodes.\par

\section{PROPOSED CORNER FEATURE REPRESENTATION AND RECOGNITION METHOD}
In our previous works, we have proposed a Boulic-sinusoidal pendulum modeling of human kinematics. The model equates the human body to six scattering centers, which can be mapped one by one into three motion patterns for all common indoor human activities. For each type of pattern, the minimum number of key corner points required to reconstruct its motion parameters can be calculated, corresponding to the feature representation with the best theoretical generalization ability. The minimum number of required corners is strictly proved to be $30$ \cite{Micro-Doppler Corner Representation}. In addition, DoG-$\mu$D-CornerDet is used for corner extraction \cite{Micro-Doppler Corner Detection}. In sequence, this section analyzes the micro-Doppler corner filter method based on polynomial fitting smoothing and maximizing the inter-class distance. Then, the extracted corners from the radar range and Doppler profiles are concatenated together into 3D point cloud. Finally, the DGNN-based decision-making model designed for 3D point cloud is proposed.\par
\subsection{Review on Definition and Extraction for Corners}
We first review the definition of micro-Doppler corner feature and the corner extraction method based on the DoG-$\mu$D-CornerDet.\par
The corner feature is obtained by mining the geometrical properties of the micro-Doppler information on $\mathbf{R^2TM}$ and $\mathbf{D^2TM}$. In the case of TWR applications, relative motion between human limb nodes induces micro-Doppler information. These features are represented on the $\mathbf{R^2TM}$ and $\mathbf{D^2TM}$ as multiple bright continuous curves corresponding to occupying range bins or Doppler bands of different widths. The stationary and inflection points on these curves, the intersections of the curves with the coordinate axes, and the intersections of the curves with each other all reflect significant gradient changes in the image. These points are defined as corners.\par
In the previous, DoG-$\mu$D-CornerDet was used to implement micro-Doppler corner point feature extraction on two types of images. The simplified flowchart is shown in Fig. \ref{DoG-mD-CornerDet}. The module is constructed by cascading the DoG supervisory end and the $\mu$D-CornerDet supervised module. The DoG module extracts pixel-level features with large gradient variations at different scales, which are used as supervisory labels for downstream $\mu$D-CornerDet training. $\mu$D-CornerDet uses ResNet-101 as backbone, upsampling and downsampling pyramids as neck, and fully-connected layers as head. The feature maps are mapped into regression and classification branches. The regression branch outputs the coordinates of the micro-Doppler corners, and the classification branch outputs whether the current micro-Doppler corner point detected is positive or negative. The network works in target detection scenario. The output Boolean value of its classification branch indicates whether the current corner point is a false alarm or not, thus reducing the noise sensitivity of the network. After training, the $\mathbf{R^2TM}$ and $\mathbf{D^2TM}$ are directly fed into $\mu$D-CornerDet, and the predict corners will be used as inputs to the following micro-Doppler filter module. \par
\subsection{Micro-Doppler Corner Representation Module}
\begin{figure}
    \centering
    \includegraphics[width=0.48\textwidth]{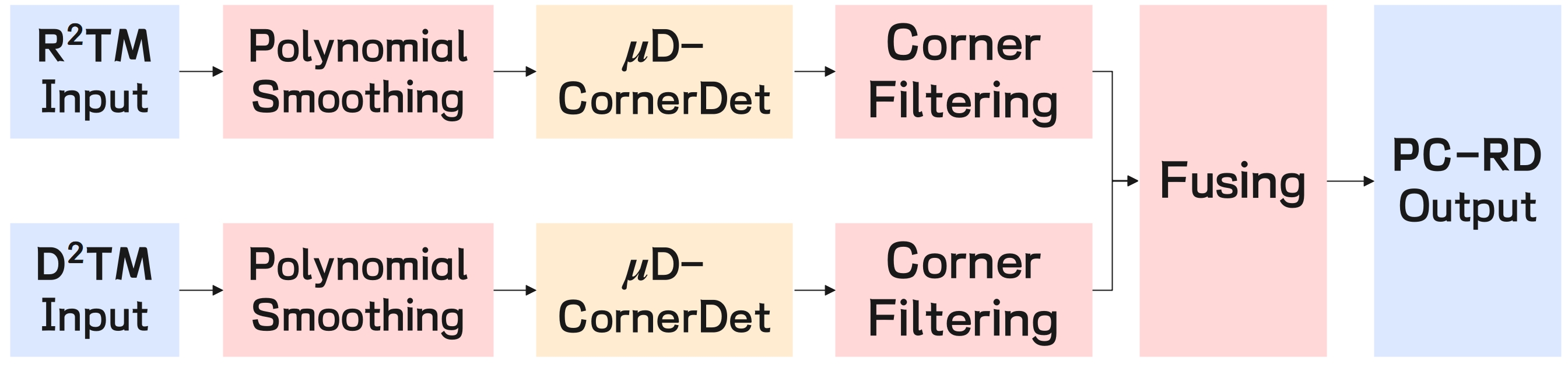}
    \caption{Scheme of the proposed micro-Doppler corner representation module.}
    \label{Micro-Doppler Corner Representation}
    \vspace{-0.0cm}
\end{figure}

In this paper, the scheme for micro-Doppler corner representation is shown in Fig. \ref{Micro-Doppler Corner Representation}, including polynomial smoothing, $\mu$D-CornerDet, and corner filtering. Polynomial smoothing is employed to make the $\mathbf{R^2TM}$ and $\mathbf{D^2TM}$ fit the kinematic model better. Unlike previous works, to ensure the fineness of image feature extraction, the number of output corners of DoG-$\mu$D-CornerDet module is $100$, rather than the required $30$ corners. In addition, micro-Doppler corner filtering is designed to ensure that the input dimensions of the back-end decision module are uniform, and maximize the inter-class distance.\par
\par
Define $\mathbf{R^2TM}$ and $\mathbf{D^2TM}$ uniformly as $I(\mathbf{px})$, where $\mathbf{px}$ denotes the horizontal and vertical axis of the input image. Expanding the input $I(\mathbf{px})$ along the horizontal direction, the unfolded vectors are sliced by using rectangular window with length of $\mathrm{Win}=2P+1,~P\in \mathbb{Z}^{+}$. In each slice, the polynomial of order $\mathrm{Ord}\in \mathbb{Z}^{+}$ is needed to fit. Here, the cost function is developed based on squared errors: 

\vspace{-0.2cm}
\begin{equation}
\delta_\mathrm{Ord}=\sum_{i=-P}^P\left(\sum_{k=0}^\mathrm{Ord} a_ki^k-x_i\right)^2,
\end{equation}
where $x_i$ is the sliced vector, $i$ is the index, $a_k,k=0,1,\ldots,\mathrm{Ord}$ is the coefficients of the polynomial. In the experiment, $\mathrm{Ord}$ is preset to $17$, which aims to ensure that the number of detectable points exceeds $30$ with certain noise robustness.\par
Then, by using the Lagrange multiplier method we can obtain $\mathrm{Win}+1$ equations \cite{Least Square Estimation}:

\vspace{-0.2cm}
\begin{equation}
\begin{gathered}
\sum_{k=0}^\mathrm{Ord} \left(\sum_{i=-P}^P i^{j+k}a_k\right)=\sum_{i=-P}^P i^j x_i\\j=0,1,\ldots,\mathrm{Ord}
\end{gathered}.
\end{equation}\par
\begin{figure}
    \centering
    \includegraphics[width=0.48\textwidth]{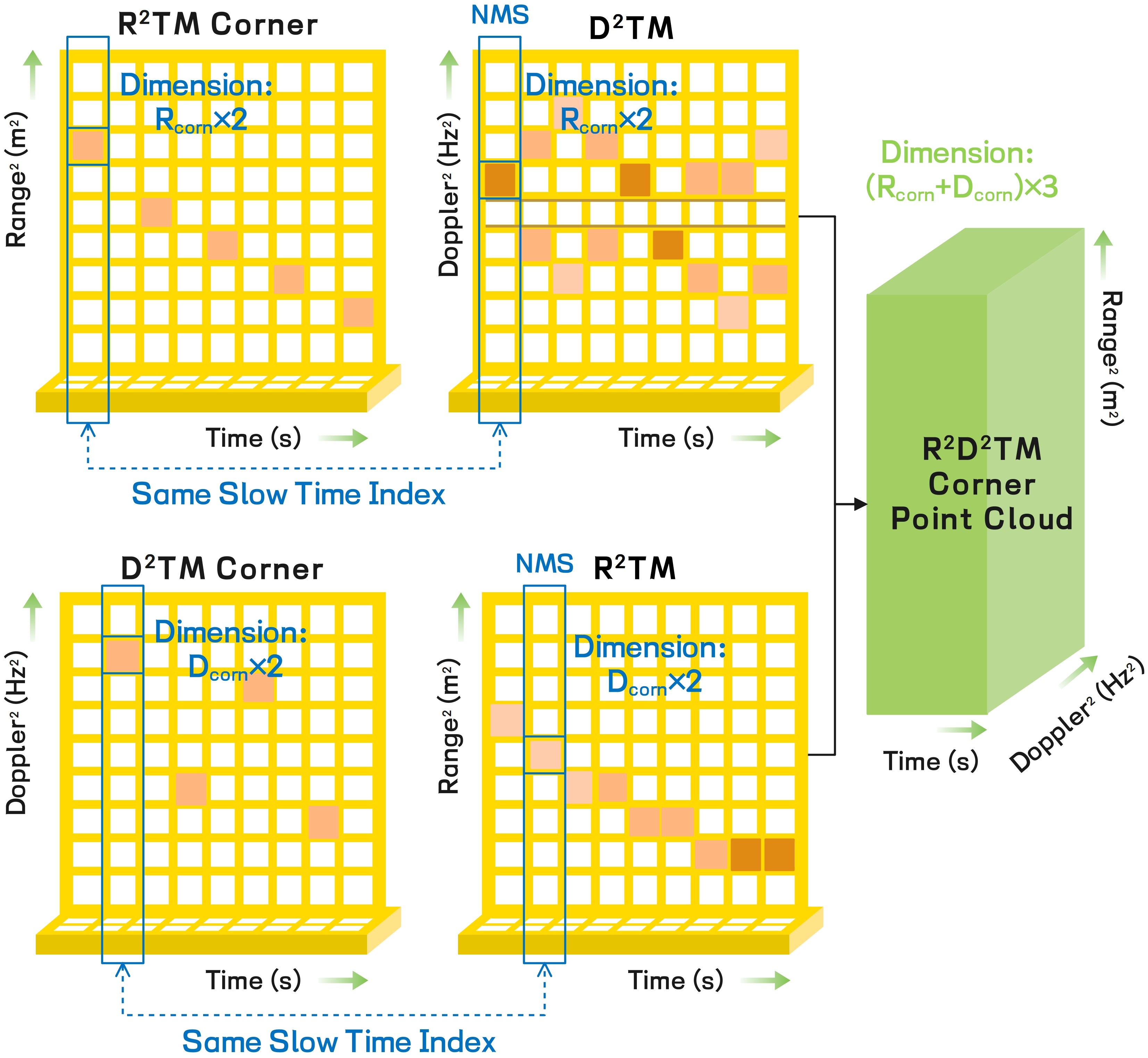}
    \caption{The process of fusing $\mathbf{R^2TM}$ and $\mathbf{D^2TM}$ corner data into $\mathbf{PC-RD}$ corner point cloud.}
    \label{Corner Fusion}
    \vspace{-0.0cm}
\end{figure}\par
The matrix form of the above equation is as follows:

\vspace{-0.2cm}
\begin{equation}
(\mathbf{A}^\top \mathbf{A})\mathbf{a}=\mathbf{A}^\top \mathbf{x}
\end{equation}
where $\mathbf{A}$ is the Vandermonde matrix with respect to the parameter $P$. The parameters are calculated using:

\vspace{-0.2cm}
\begin{equation}
\mathbf{a}=\mathbf{A}^{+}\mathbf{x},
\end{equation}
where:

\vspace{-0.2cm}
\begin{equation}
\mathbf{a}=[a_{0},a_{1},\ldots,a_{n}]^{\top},
\end{equation}\par
\vspace{-0.2cm}
\begin{equation}
\mathbf{x}=[x_{-P},\ldots,x_{-1},x_{0},x_{1},\ldots,x_{P}]^{\top},
\end{equation}\par
\vspace{-0.2cm}
\begin{equation}
\mathbf{A}=\begin{bmatrix}1&-P&(-P)^2&\cdots&(-P)^n\\\vdots&\vdots&\vdots&&\vdots\\1&0&0&\cdots&0\\\vdots&\vdots&\vdots&&\vdots\\1&P&(P)^2&\cdots&(P)^n\end{bmatrix},
\end{equation}\par
\vspace{-0.2cm}
\begin{equation}
\mathbf{A^{+}=(A^{\top}A)^{-1}A^{\top}}.
\end{equation}\par
After getting the fitting parameters, the original signal $x_i$ in each slice is replaced with the smoothed polynomial curve $\mathrm{Curve}_i=\sum_{k=0}^\mathrm{Ord} a_k i^k$. All the processed slices are recovered back to the scale of the original image, which are fed into $\mu$D-CornerDet. When the number of corners is not sufficient for the model, the window length is increased until the number of points meets the requirement $\mathrm{Cor_0}$.\par
\begin{algorithm}[!t]
\DontPrintSemicolon
  \KwInput{$\mathbf{I}=\mathbf{R^2TM}$ or $\mathbf{I}=\mathbf{D^2TM}$}
  \KwOutput{Corresponding filtered corner feature $\mathbf{Cor}_\mathrm{out}$.}
  Initializing $\mathrm{Init}=2$, $\mathbf{Cor}_\mathrm{out}=\mathrm{zeros}(\mathrm{Cor_0},2)$;\;
  \tcc{$\mathrm{Cor_0}$ is the output number of corners in each map ($30$ in practice).}
  \While{$\mathrm{True}$}{
    $\mathbf{Cor}=\mu\text{D-CornerDet}(\text{Polynomial-Fitted}(\mathbf{I}))$;\;
    Randomly select $1$ of the $\mathrm{Cor}_0-\mathrm{Init}$ points in $\mathbf{Cor}$ and compute $r$;\;
    Iterate over the remaining points and compute $r^{\prime}$, which is added to the set of corner points $\mathbf{Cor}_\mathrm{out}$ if $r^{\prime}\geq r$;\;
    Reset $r=r^\prime$, $\mathrm{Init}=\mathrm{Init}+1$, until $\mathrm{Init}>\mathrm{Cor_0}$ to stop looping.\;
  }
\caption{Method of Corner Filtering}
\label{Corner Filtering}
\end{algorithm}\par
Here we propose to apply a corner point screening method based on T-distributed stochastic proximity embedding (T-SNE) analysis.\par
Firstly, $\mathrm{Bat}$ groups of data are randomly selected, and the $\mathrm{Cor}$ points obtained from each group are formed into high dimensional vectors $\mathbf{u}_1,\mathbf{u}_2,\ldots,\mathbf{u}_\mathrm{Bat}$. Each vector corresponds to a set of coordinates of all corner points detected on the range and Doppler profiles. All vectors correspond to the randomly selected $\mathrm{Bat}$ groups from the data set. The high dimensional vectors are used to measure similarity between vectors using Gaussian kernel distance:

\vspace{-0.2cm}
\begin{equation}
p_{j|i}=\frac{\exp(-\parallel \mathbf{u}_i-\mathbf{u}_j\parallel^2/2\sigma^2)}{\sum_{l\neq i}\exp(-\parallel \mathbf{u}_i-\mathbf{u}_l\parallel^2/2\sigma^2)},
\end{equation}
where $l=1,2,\ldots,i-1,i+1,\ldots,\mathrm{Bat}$, and $\sigma=2^{1/k}$, where $k$ is kept the same as the total number of scaling scales in DoG. Further, the correlation probability between two data sets is defined as $p_{ij}=\frac{p_{j|i}+p_{i|j}}{2\cdot\mathrm{Bat}}$.\par
For the selected $\mathrm{Bat}$ groups of data sets, two-dimensional (2D) feature maps $\mathbf{y}_1,\mathbf{y}_2,\ldots,\mathbf{y}_\mathrm{Bat}$ are randomly generated in the 2D plane. Similarly, the Euclidean distance is utilized to define:

\vspace{-0.2cm}
\begin{equation}
q_{ij}=\frac{\left(1+\parallel \mathbf{y}_i-\mathbf{y}_j\parallel^2\right)^{-1}}{\sum_l\sum_{l^{\prime}\neq l}(1+\parallel \mathbf{y}_l-\mathbf{y}_{l^{\prime}}\parallel^2)^{-1}},
\end{equation}
where $l,~l^{\prime}$ are variables used to traverse the input vectors.\par
Finally, the optimal corners required can be obtained by finding the minimum value of the Kullback-Leibler (KL) divergence iteratively:

\vspace{-0.2cm}
\begin{equation}
\arg\min\text{KL}(P||Q)=\sum_{i\neq\mathbf{j}}p_{ij}\cdot\log(\frac{p_{ij}}{q_{ij}}).
\end{equation}\par
The solution of this optimization problem is the value of each $\mathbf{y}_1,\mathbf{y}_2,\ldots,\mathbf{y}_\mathrm{Bat}$, and it is sufficiently calculated using gradient descent. Find the center $\mathbf{y}_s$ and the maximum Euclidean distance $r$ from the center to all the projected data on the 2D plane:

\vspace{-0.2cm}
\begin{equation}
\mathbf{y}_s=\frac{1}{\mathrm{Bat}}\sum_{i=1}^\mathrm{Bat}\mathbf{y}_i,
\end{equation}
\vspace{-0.2cm}
\begin{equation}
r=\max_{\mathbf{y}_l}\|\mathbf{y}_l-\mathbf{y}_s\|^2,
\end{equation}
where $l=1,2,\ldots,y_\mathrm{Bat}$ is the index. The steps of corner filtering are given by Algorithm \ref{Corner Filtering}. In this way, the required $30$ corners $\mathbf{PC-R}_{(30\times 2)}$ and $\mathbf{PC-D}_{(30\times 2)}$ are obtained for both $\mathbf{R^2TM}$ and $\mathbf{D^2TM}$, respectively, which characterize human activity pattern and fit the kinematic motion model.\par
The output of the micro-Doppler corner representation module is a three-dimensional point cloud, which is obtained by fusing $\mathbf{PC-R}$ and $\mathbf{PC-D}$. The process is shown in Fig. \ref{Corner Fusion}. For each corner point in $\mathbf{PC-R}$, the point on the $\mathbf{D^2TM}$ corresponding to the same slow time is determined by non-maximal suppression (NMS) along frequency dimension \cite{NMS}. Similarly for each corner point in $\mathbf{PC-D}$, the point on the $\mathbf{R^2TM}$ corresponding to the same slow time is determined by NMS along range dimension. The final obtained $\mathbf{PC-RD}$ is a $\mathrm{Cor}\times 3$ three-dimensional point cloud with the size of $60\times 3$, which is used as the training and inference input for the subsequent DGNN-based decision module.\par
\subsection{Linked Dynamic Graph Learning Module}
\begin{figure*}
    \centering
    \includegraphics[width=\textwidth]{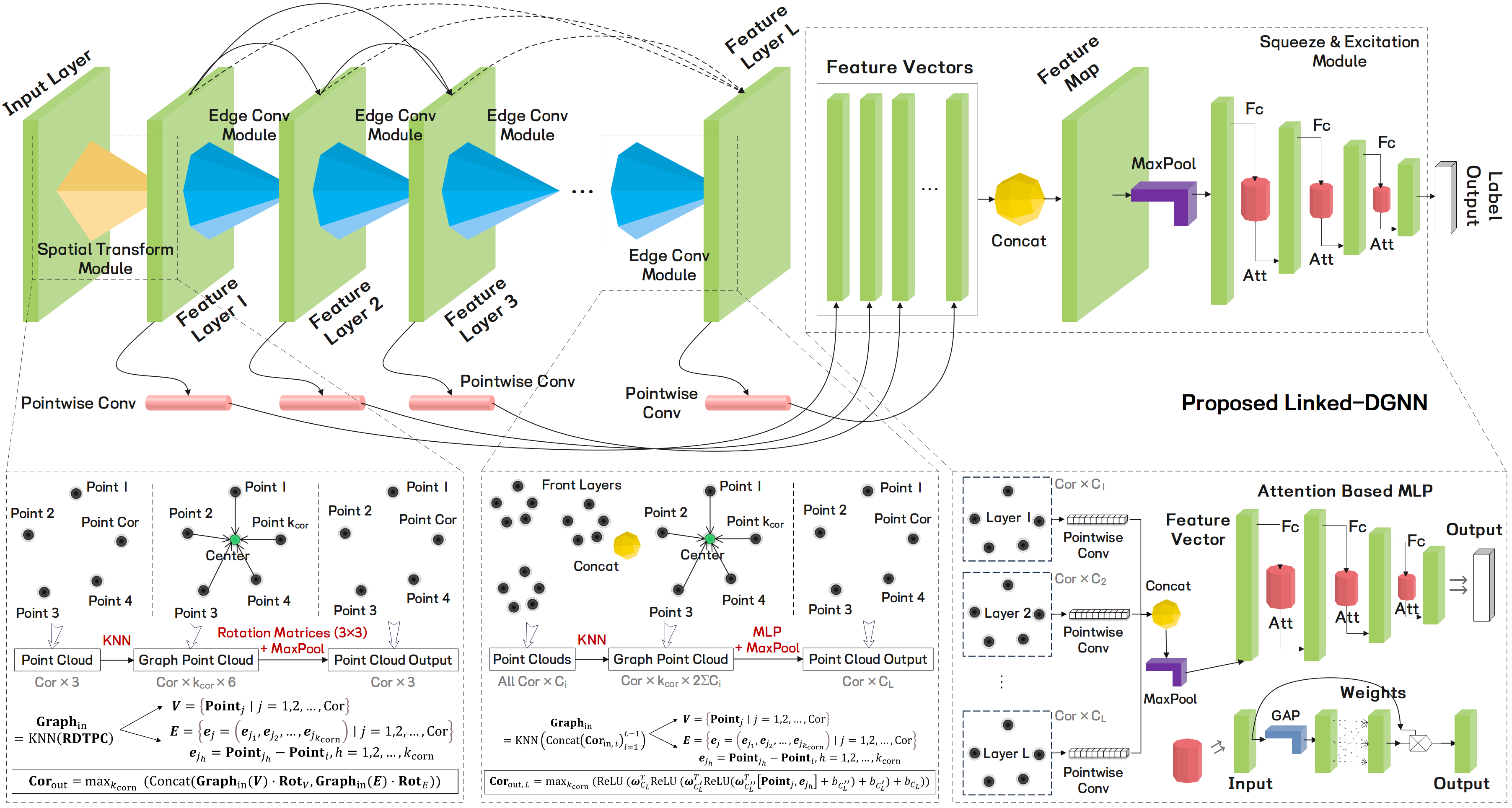}
    \caption{Block diagram of the proposed DGNN-based method, containing the design details of the graph conducting, graph focusing and graph decision modules in the network.}
    \label{Linked DGNN}
    \vspace{-0.0cm}
\end{figure*}

Fig. \ref{Linked DGNN} illustrates the overall structure of the recognition decision network \cite{Linked-DGNN 1, Linked-DGNN 2}. The structure not only considers the sparsity of the micro-Doppler corner point cloud, but also introduces the concept of multi-scale feature pyramid, which fuses the point clouds from different feature extraction layers to make final decision.\par
The designed recognition network contains spatial transform module, edge convolution module, and squeeze and excitation module. In the following, we describe the design principles for these three modules in detail.\par
\par
The spatial transform module is developed for graph conducting, which maps the input three-dimensional point cloud to a directed point cloud graph. Here, we use the K-Nearest-Neighbors (KNN) algorithm to cluster this $60\times 3$ point cloud, $k_\mathrm{cor}$ nearest neighbor sample points are clustered in each class. All the points within a cluster are created as edges pointing to the cluster center, which should also be represented as a three-dimensional coordinate that equals to the difference between the three-dimensional coordinate of the current point and the cluster center. By adding all the 3D coordinate information of the edges obtained above to the input three-dimensional point cloud, the coordinate information of each point should be changed from $3$ dimensions to $6$ dimensions. Then, we introduce the multiplication of two $3 \times 3$ matrices, and all elements are set as learnable weights. The two matrices are used to learn the rotation angles or scaling scales that may exist spatially for different input point cloud data. Finally, the dimension $k_\mathrm{cor}$ is pooled off by a maximum pooling layer to output an $60\times 3$ point cloud with exactly the same dimension as the input.\par
\begin{table}
\begin{center}
\caption{Uniform Parameters of Simulated and Measured TWR System.\label{System Parameters}}
\resizebox{0.48\textwidth}{!}{
\begin{tabular}{cc}
\hline\hline
\textbf{Parameters}             & \textbf{Value}     \\ 
\hline
Antenna Transceiver Spacing     & (SISO) $0.15 \mathrm{~m}$   \\
Work Center Frequency          & $1.5 \mathrm{~GHz}$   \\
Band Width & $2.0 \mathrm{~GHz}$\\
Sampling Points & $1024$                \\
Sampling Period & $4 \mathrm{~s}$                 \\
Wall Thickness & $0.12 \mathrm{~m}$  \\
Human Motion Range from Radar & $1 \sim 4 \mathrm{~m}$     \\
SNR of Raw Data$^{1}$ & $-19.85 \sim -12.46 \mathrm{~dB}$ \\
SNR of Processed Data$^{1}$ & $\approx0 \mathrm{~dB}$ \\
Antenna Height to Ground & $1.5 \mathrm{~m}$      \\ 
Number of Activities & $12$ \\
Training Set & $3200$ \\
Validation Set$^{2}$ & $800$ \\
Test Set$^{2}$ & $400$ for Each Tester \\
\hline\hline
\end{tabular}
}
\end{center}
\footnotesize $^{1}$ Approximate estimates are obtained from multiple sets of data collected from a single hollow block wall (Relative dielectric $\varepsilon_r \approx 6$). The SNR mentioned in the table are obtained by manually selecting the target's region of the image and calculating the image pixel energy.\\
\footnotesize $^{2}$ As the core of this work is to validate the generalization of the proposed method, therefore, the test sets and the validation set are collected separately. The ratio of the amount of data in the training set, validation set and each test set is controlled as $8:2:1$.\\
\vspace{-0.0cm}
\end{table}\par
The edge convolution module is developed for graph focusing, which is used to process the edge matrix on the point cloud graph representation. Assuming that the current edge convolution module is the network's $L^\mathrm{th}$ layer, then its input is the concatenation of all the output point clouds from $1$ to $L-1$ layers. Similar to the KNN operation in spatial transform module, the final obtained graph should be a $\mathrm{Cor}\times k_\mathrm{cor} \times 2\sum_{i=1}^{L-1}C_i$ dimensional matrix. Then, MLP is used to extract the information in this large matrix, and the number of nodes in the three layers of MLP are $2\sum_{i=1}^{L-1}C_i,~\sum_{i=1}^{L-1}C_i,C_L$, and ReLU is chosen for the activation function between the layers. Finally, followed by the maximum pooling layer, the output dimension of edge convolution module is a $60\times C_L$ point cloud.\par
\begin{figure*}
    \centering
    \includegraphics[width=\textwidth]{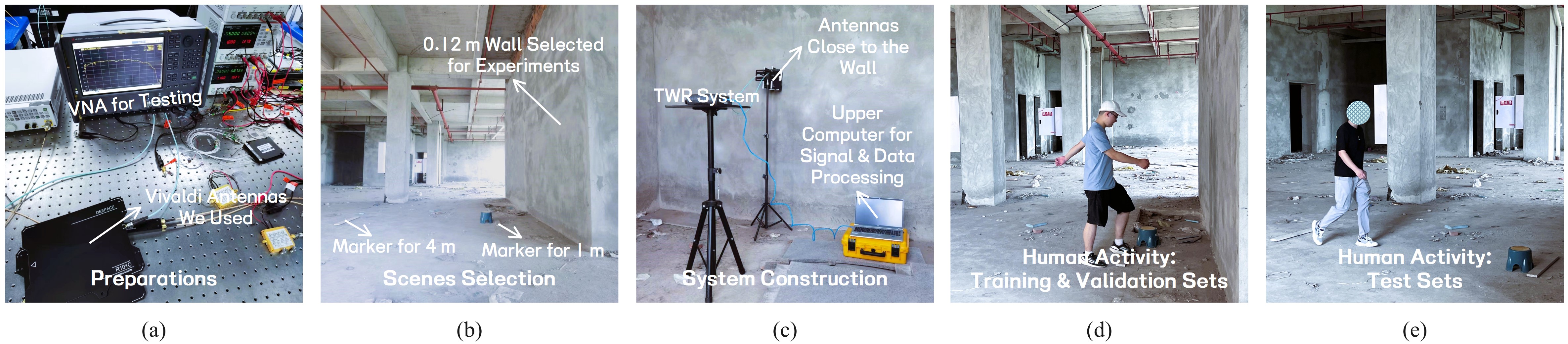}
    \caption{Field photos of the experiments: (a) Equipment preparations, (b) Scenes selection, (c) TWR system construction, (d) Acquisition of human activity samples for training and validation sets, (e) Acquisition of human activity samples for test sets.}
    \label{Experiment Photos}
    \vspace{-0.0cm}
\end{figure*}\par
The squeeze and excitation module is constructed by a one-dimensional sequence version of attention based MLP \cite{MLP}, designed for graph decision making. The outputs of all the $L$ layers of the edge convolution module are $\mathrm{Cor}\times C_1$ to $\mathrm{Cor}\times C_L$ point clouds, which are then concatenated and downsampled into a $1 \times 1024$ sequence using pointwise convolution and maximum pooling. Then it is mapped as recognition labels using MLP with node counts of $512$, $256$, and $12$, respectively. The vectors are stretched between layers into $8\times 8 \times 16$, $8 \times 8 \times 8$, and $8 \times 8 \times 4$ matrices, processed with the channel attention mechanism \cite{Channel Attention}, and re-stretched into vectors of lengths $1024$, $512$, and $256$. The main purpose of this step is to alleviate the problem of mutual interference of unrelated semantic information between layers.\par
The loss function of the network is developed using cross entropy \cite{Cross Entropy}, which can be expressed as:

\vspace{-0.2cm}
\begin{equation}
\mathrm{Loss}=-\frac{1}{\mathrm{Bat^{\prime}}}\sum_{\mathrm{bat^{\prime}}=1}^{\mathrm{Bat^{\prime}}}\sum_{\mathrm{cla}=1}^{12}\mathrm{lab}_\mathrm{bat^{\prime}}^\mathrm{cla}\mathrm{log}(\widetilde{\mathrm{lab}}_\mathrm{bat^{\prime}}^\mathrm{cla})
\end{equation}
where $\mathrm{Bat^{\prime}}$ is the batch size, and $\mathrm{cla}$ is the index of activity classes. $\mathrm{lab}_\mathrm{bat^{\prime}}^\mathrm{cla}$ is the ground-truth and $\widetilde{\mathrm{lab}}_\mathrm{bat^{\prime}}^\mathrm{cla}$ is the network prediction.\par

\section{EXPERIMENTAL VERIFICATION}
In this section, numerical simulated and measured experiments are carried out to verify the performance of the proposed method. First, the developed TWR system and the data set are introduced. Second, the performance is verified by using visualization and confusion matrix. Next, detailed comparison and ablation experiments are conducted in terms of accuracy and generalization ability.\par
\subsection{System Design and Data Collection}
As shown in TABLE \ref{System Parameters}, the parameters used in simulated and measured TWR system are presented. The simulated data sets are referenced from open-source results from University College London (UCL) \cite{UCL}, and the measurements are captured by the developed prototype.\par
In the experiment, the TWR system transmits and receives UWB waveforms. The transceiver antennas are spaced $0.15~m$ apart and are $1.5~m$ from the ground. The total number of samples along fast time and slow time is $1024$, where fast time axis represents range bins from $0\sim 4~m$ and slow time axis represents time windows from $0\sim 4~s$. A single hollow brick wall with thickness $0.12~m$ is used. In the simulation, an $3\times 0.12 \times 2~m$ isotropic rectangular medium is used. The relative dielectric constant is set to $6$.\par
\begin{table}
\begin{center}
\caption{Hyperparameters for Network Training and Validation.\label{Training Settings}}
\resizebox{0.48\textwidth}{!}{
\begin{tabular}{cc}
\hline\hline
\textbf{Network Hyperparameters}             & \textbf{Value}          \\ \hline
Batch Size                      & $32$                     \\
Total Epoches                   & $20$                    \\
Initial Learning Rate           & $0.00147$                   \\
Regulization Method             & $L - 2$           \\
Validation Frequency (Batches)$^{1}$  & $10$                       \\
Optimizer                       & Adam  \\
Batch Normalization Statistics  & Population              \\
Training Hardware               & NVIDIA Tesla V$100$, $16~\mathrm{G}$  \\
Validation Hardware             & NVIDIA RTX $3060$, $12~\mathrm{G}$, OC   \\
Training and Validation Software  & Python $3.8$, Paddle $2.4$   \\
\hline\hline
\end{tabular}
}
\end{center}
\footnotesize $^{1}$ Each epoch contains $100$ batches, for a total of $20$ epochs, the total number of validations is $200$.\\
\vspace{-0.0cm}
\end{table}\par
The designed identification labels contained $11$ human activities and $1$ empty scene, i.e. $12$ categories, including: $S1$, Empty; $S2$, Punching; $S3$, Kicking; $S4$, Grabbing; $S5$, Sitting Down; $S6$, Standing Up; $S7$, Rotating; $S8$, Walking; $S9$, Sitting to Walking; $S10$, Walking to Sitting; $S11$, Falling to Walking; $S12$, Walking to Falling. The training and validation sets are collected based on tester with height of $1.8~m$. For each human activity $S2 \sim S12$, $300$ samples are collected. For empty scene $S1$, $700$ samples are collected. In general, the data set is divided randomly as $80\%$ samples for training and $20\%$ samples for validation. By contrast, testers with different heights are employed to generate the test data sets. In the simulation experiments, the heights for testers are $1.8$, $1.7$, $1.6$, and $1.5$ meters while in the measurement, the heights are $1.8$, $1.7$, and $1.6$ meters. For each height, $400$ samples are collected. Experimental photos are shown in Fig. \ref{Experiment Photos}. \par
As shown in TABLE \ref{Training Settings}, some hyperparameters are given. The batch size for network training is set to $32$. A total of $2000$ batches for training and $200$ batches for validation. Adam optimizer is used with an initial learning rate of $0.00147$. The hardware condition for training is a single NVIDIA Tesla $\mathrm{V}100$ GPU single card and the framework is the Paddle $2.4$ for python $3.8$.\par
\begin{figure*}[!ht]
    \centering
    \includegraphics[width = \textwidth]{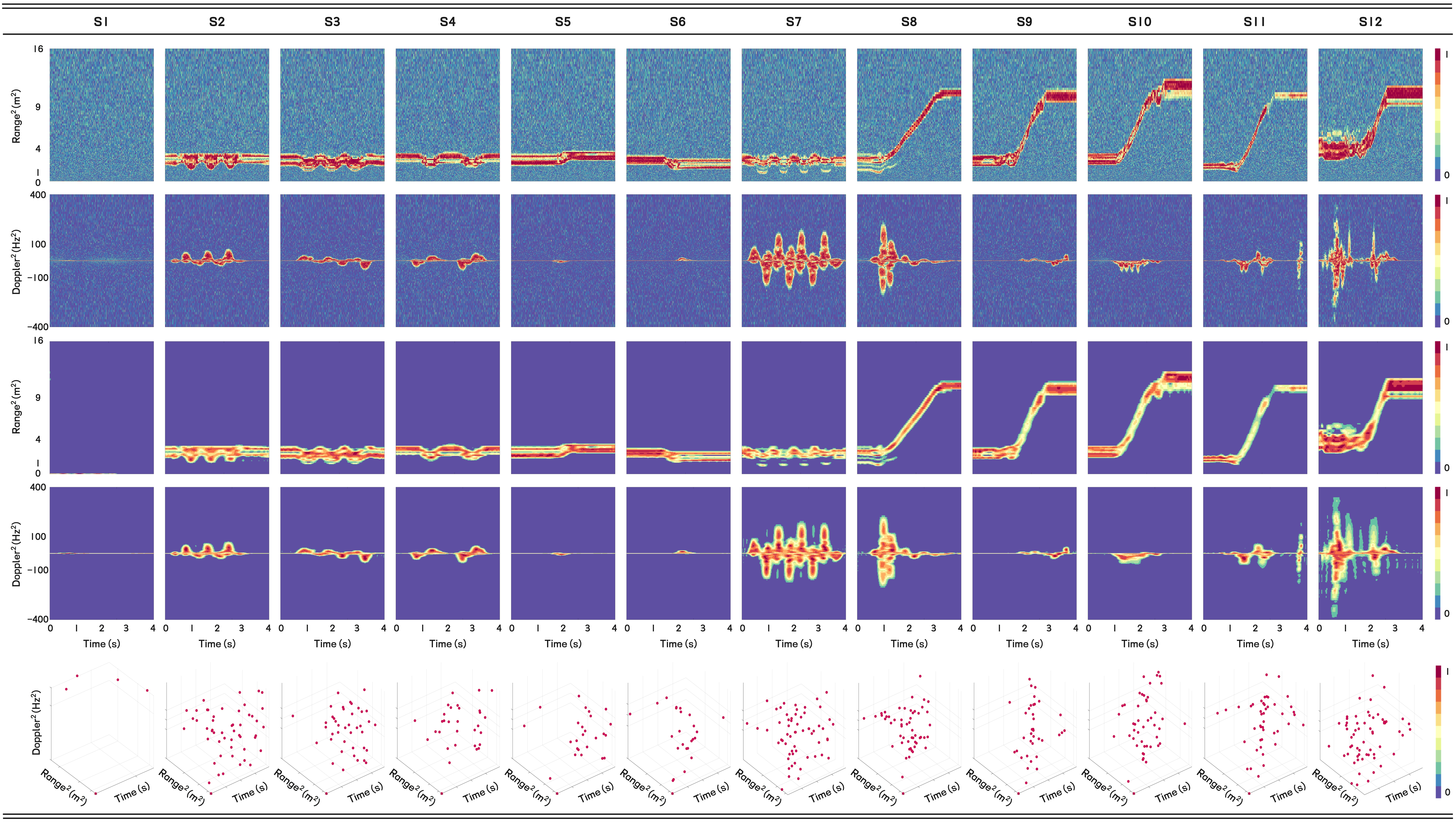}
    \caption{Visualization of the simulated data: The first and second rows are the $\mathbf{R^2TM}$ and $\mathbf{D^2TM}$, the third and fourth rows are the polynomial fit smoothed $\mathbf{R^2TM}$ and $\mathbf{D^2TM}$, respectively, and the fifth row is the $\mathbf{PC-RD}$ point cloud after corner filtering and fusion.}
    \label{Simulated Visualization}
    \vspace{-0.0cm}
\end{figure*}\par
\begin{figure*}[!ht]
    \centering
    \includegraphics[width = \textwidth]{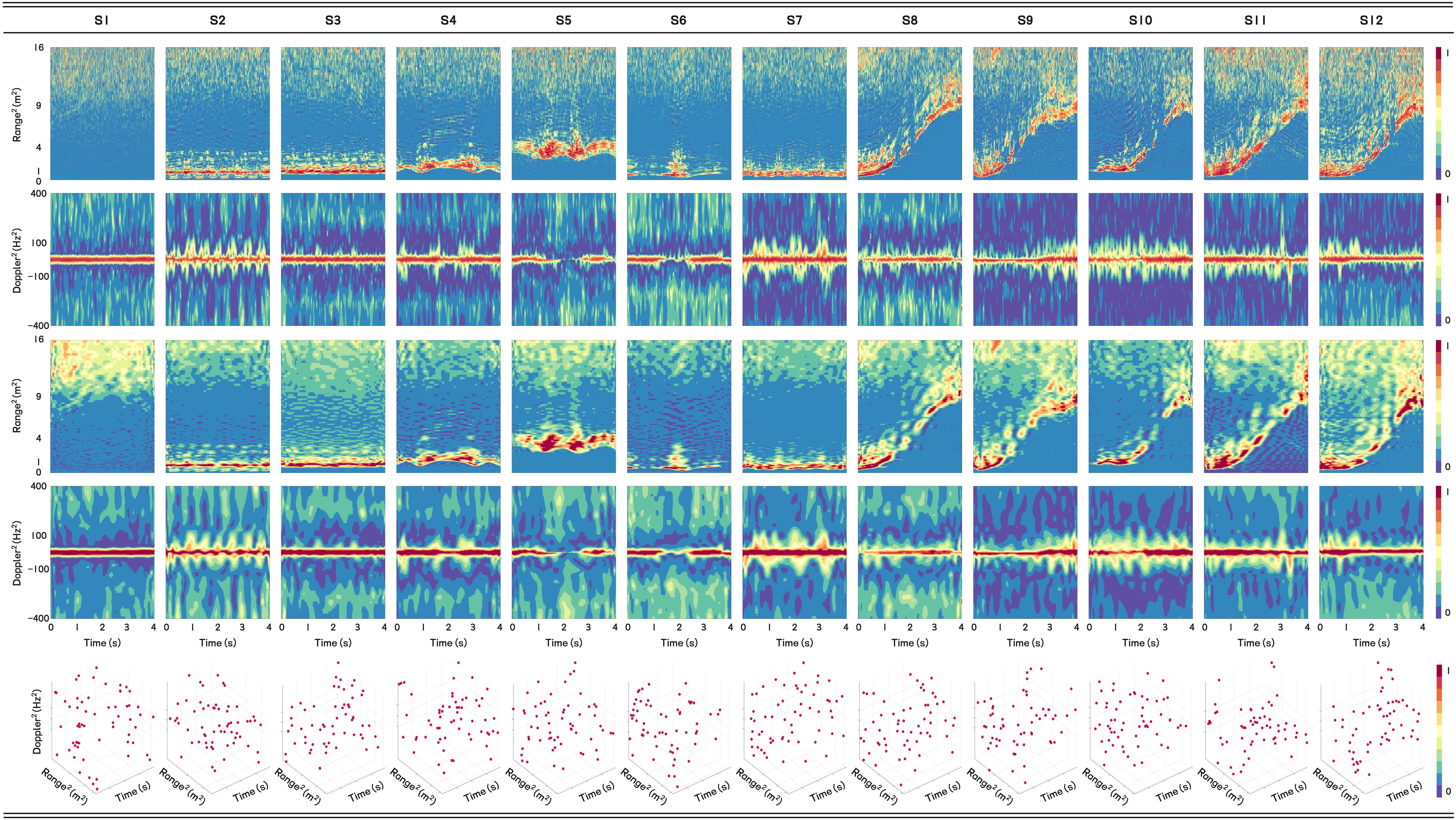}
    \caption{Visualization of the measured data: The first and second rows are the $\mathbf{R^2TM}$ and $\mathbf{D^2TM}$, the third and fourth rows are the polynomial fit smoothed $\mathbf{R^2TM}$ and $\mathbf{D^2TM}$, respectively, and the fifth row is the $\mathbf{PC-RD}$ point cloud after corner filtering and fusion.}
    \label{Measured Visualization}
    \vspace{-0.0cm}
\end{figure*}\par
\begin{figure*}[!ht]
    \centering
    \subfigure[]{\includegraphics[width=0.328\textwidth]{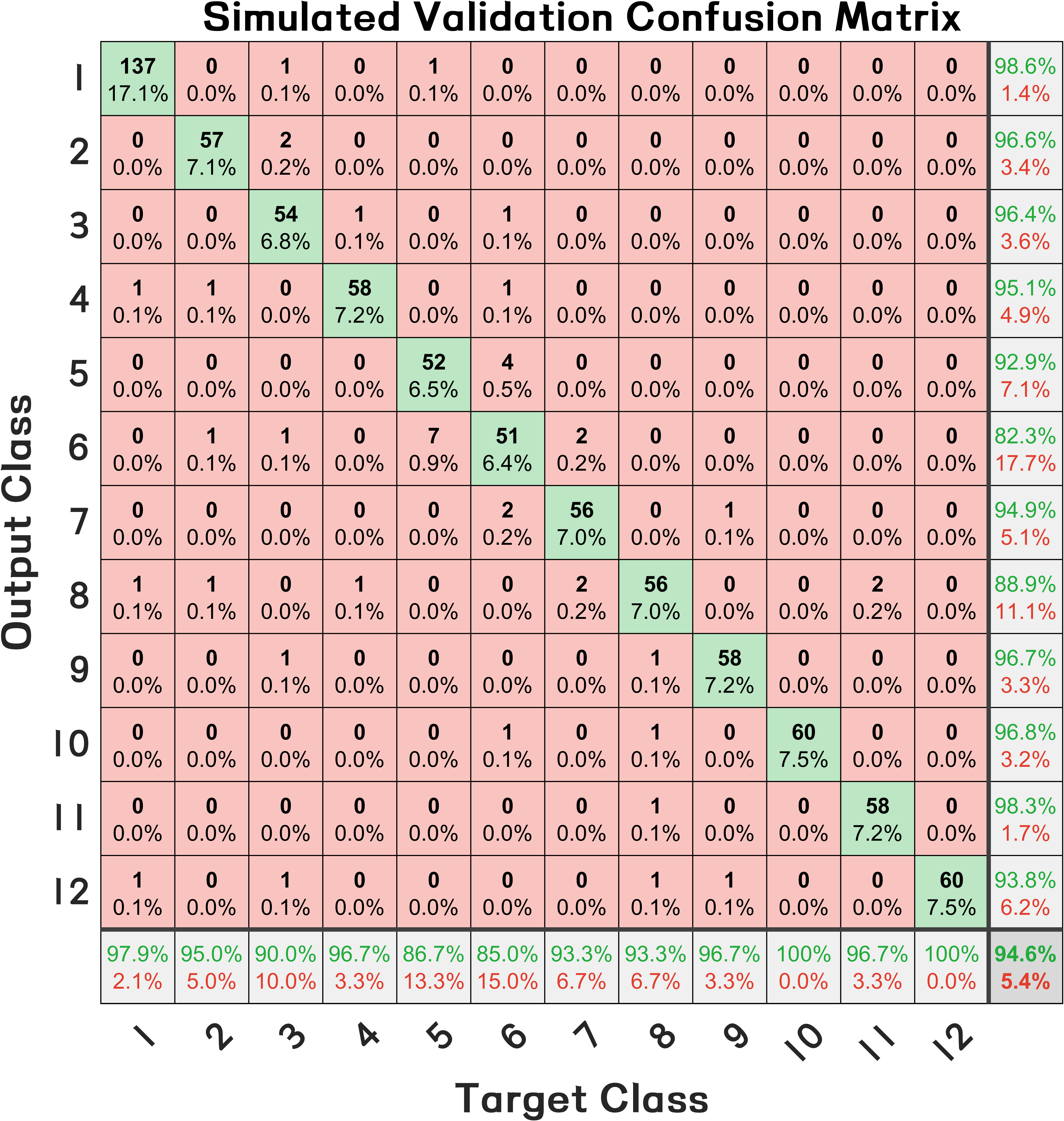}}
    \subfigure[]{\includegraphics[width=0.328\textwidth]{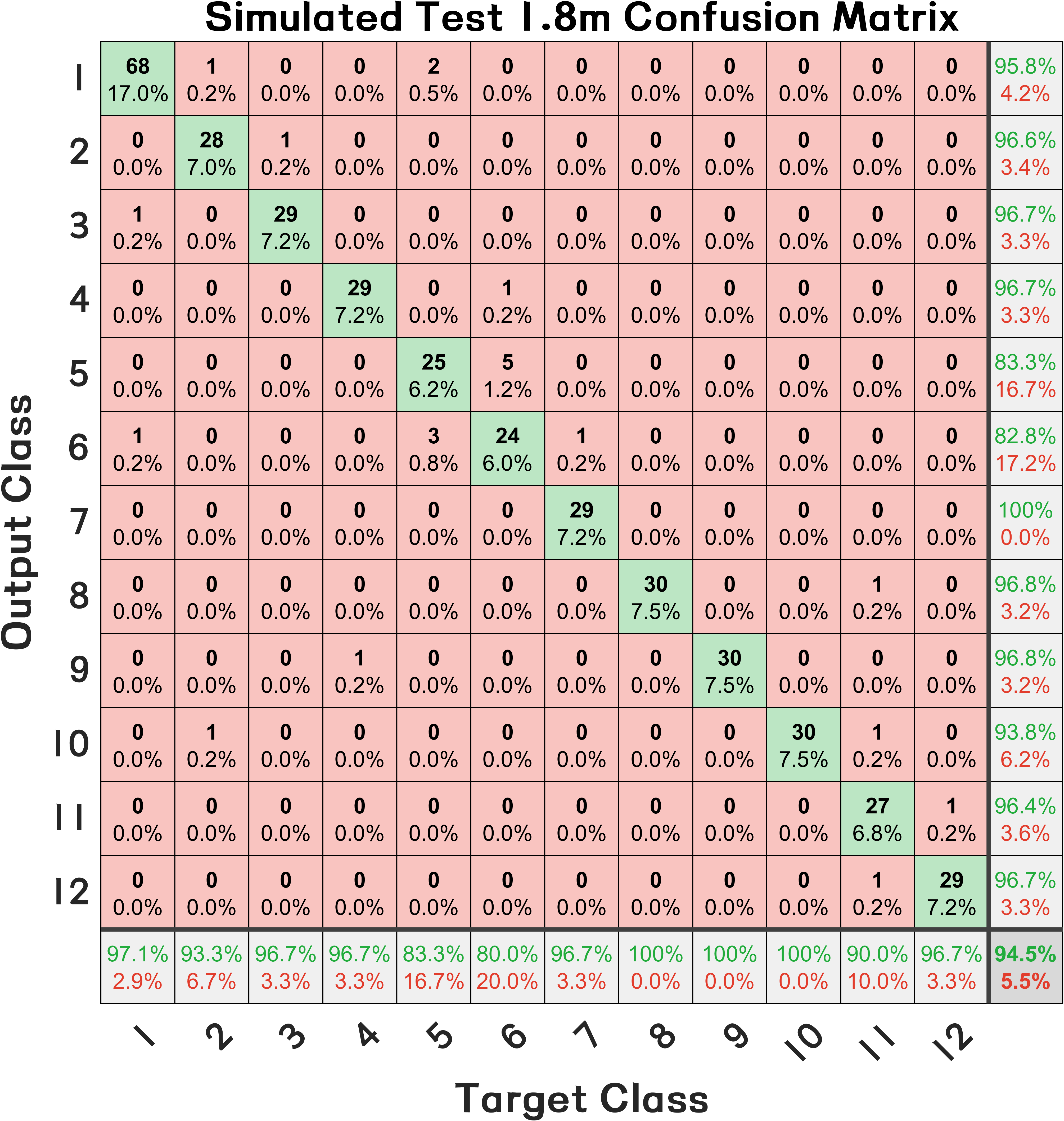}}
    \subfigure[]{\includegraphics[width=0.328\textwidth]{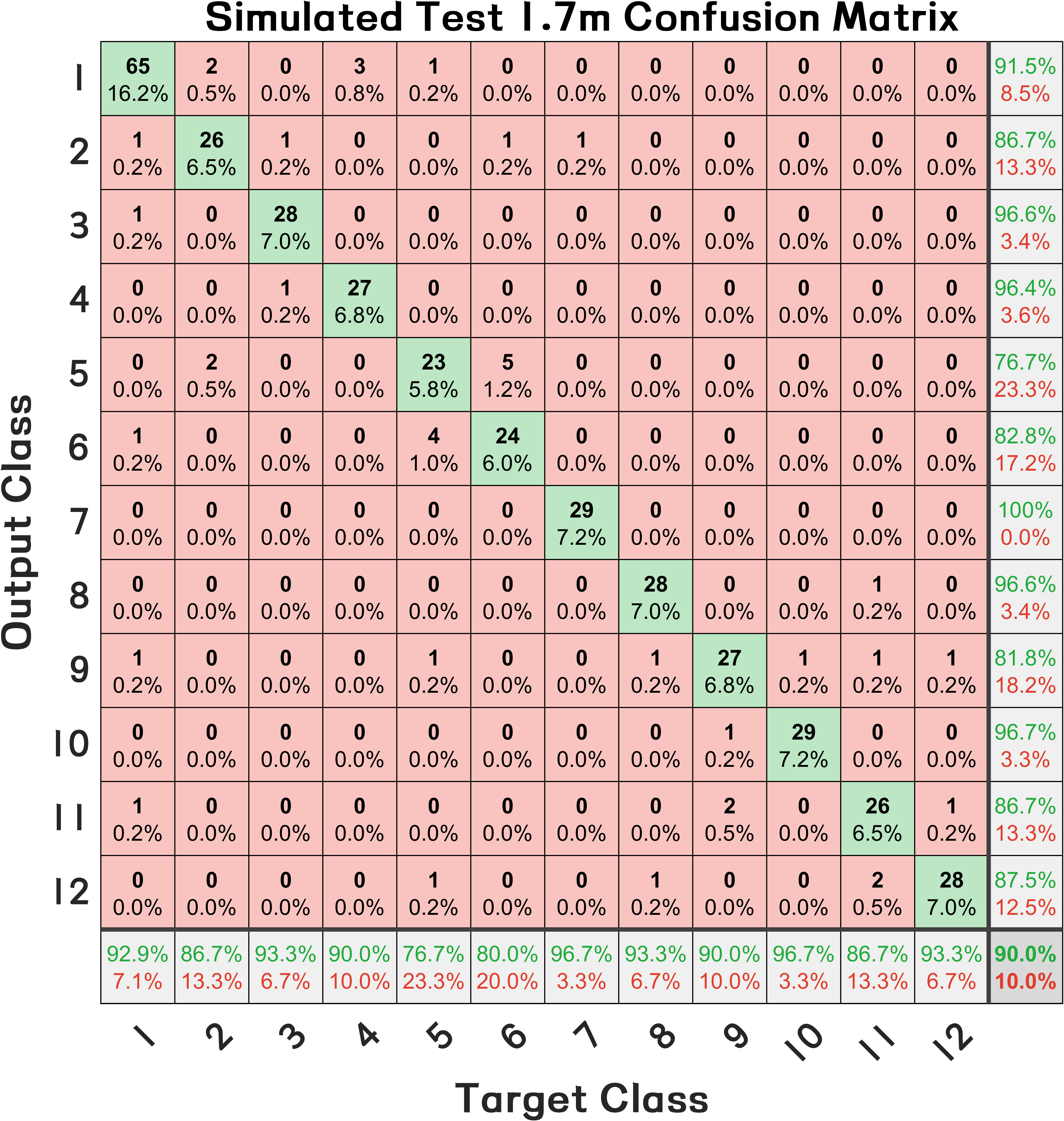}}
    \subfigure[]{\includegraphics[width=0.328\textwidth]{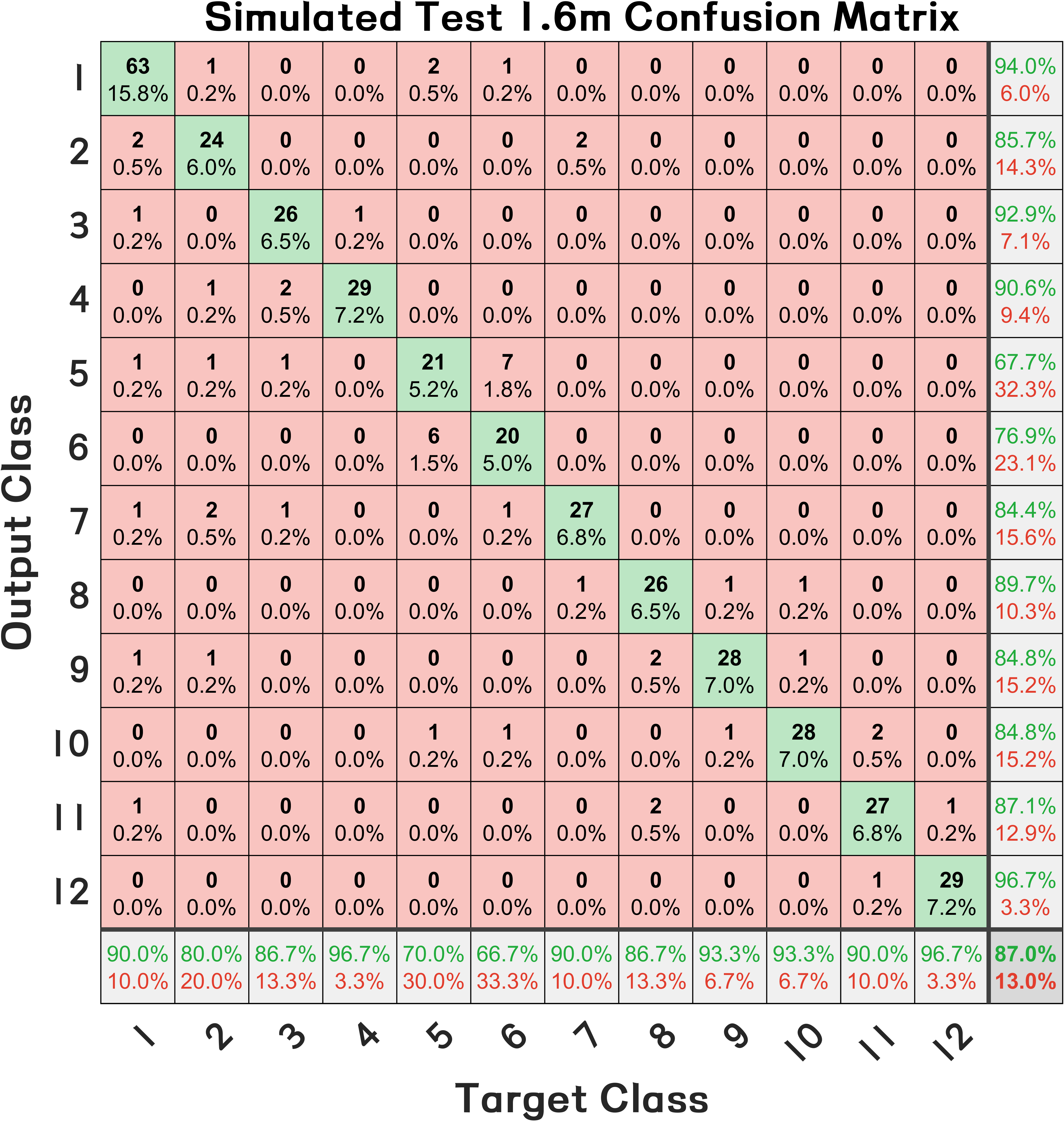}}
    \subfigure[]{\includegraphics[width=0.328\textwidth]{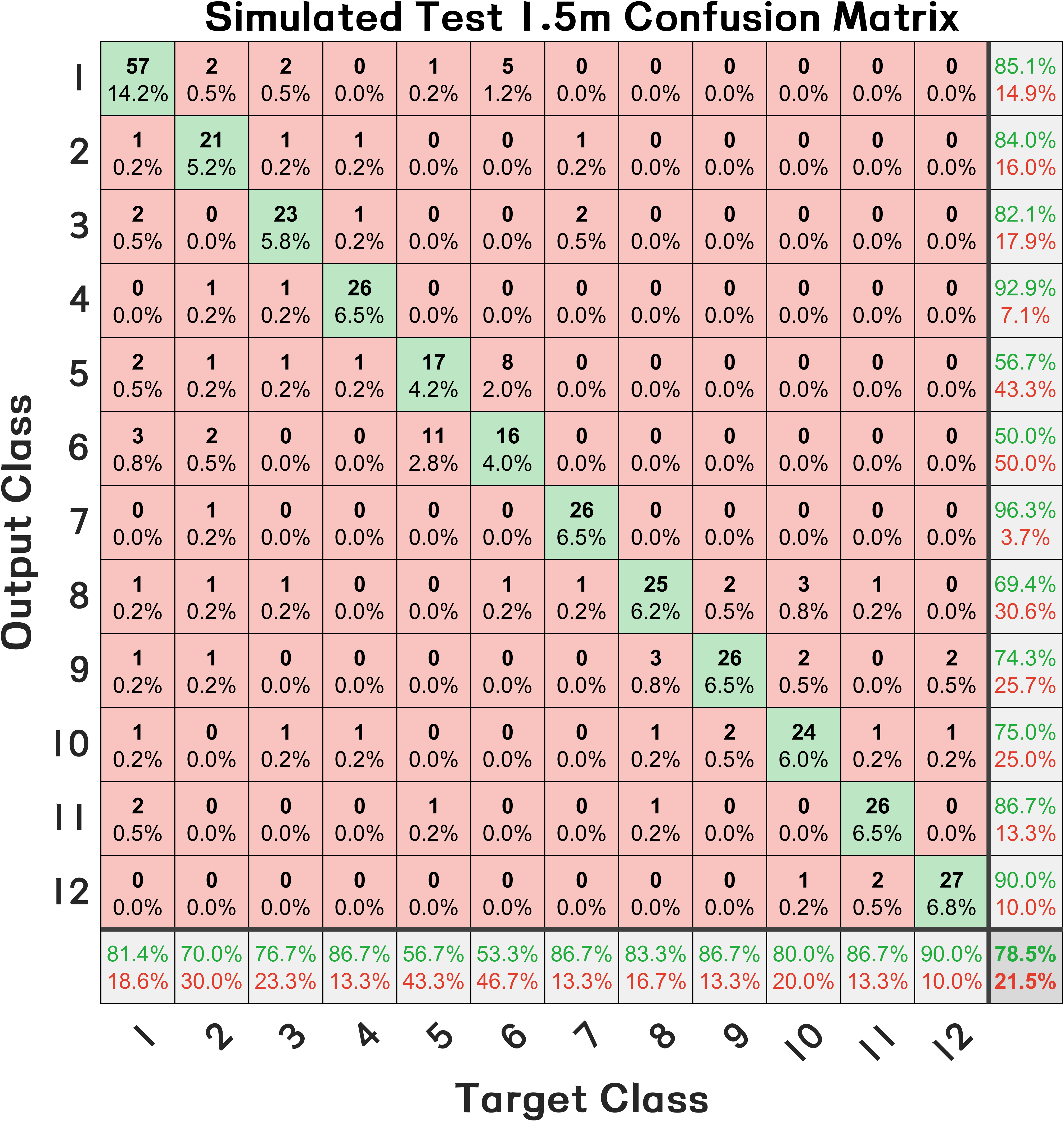}}
    \subfigure[]{\includegraphics[width=0.328\textwidth]{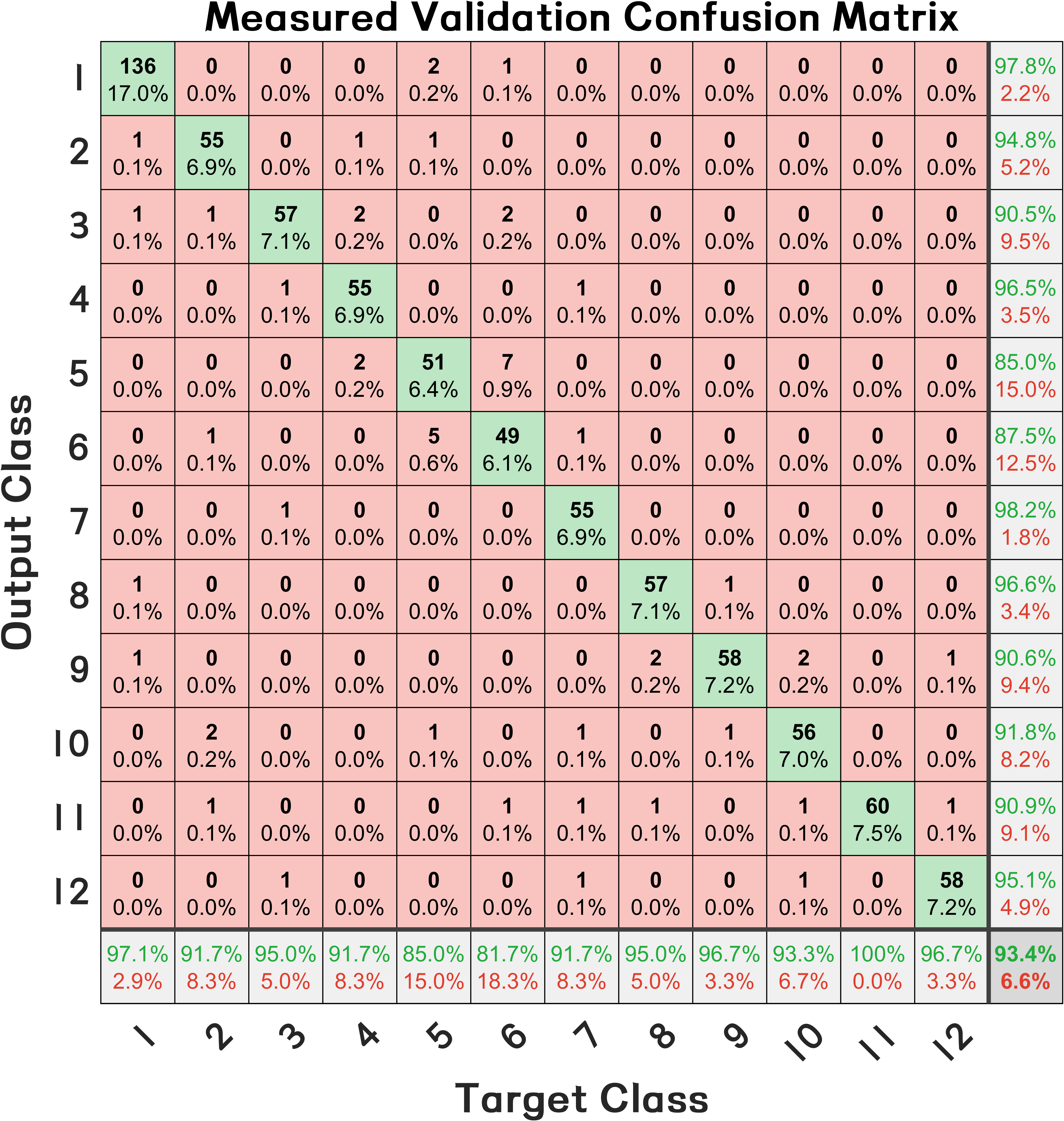}}
    \subfigure[]{\includegraphics[width=0.328\textwidth]{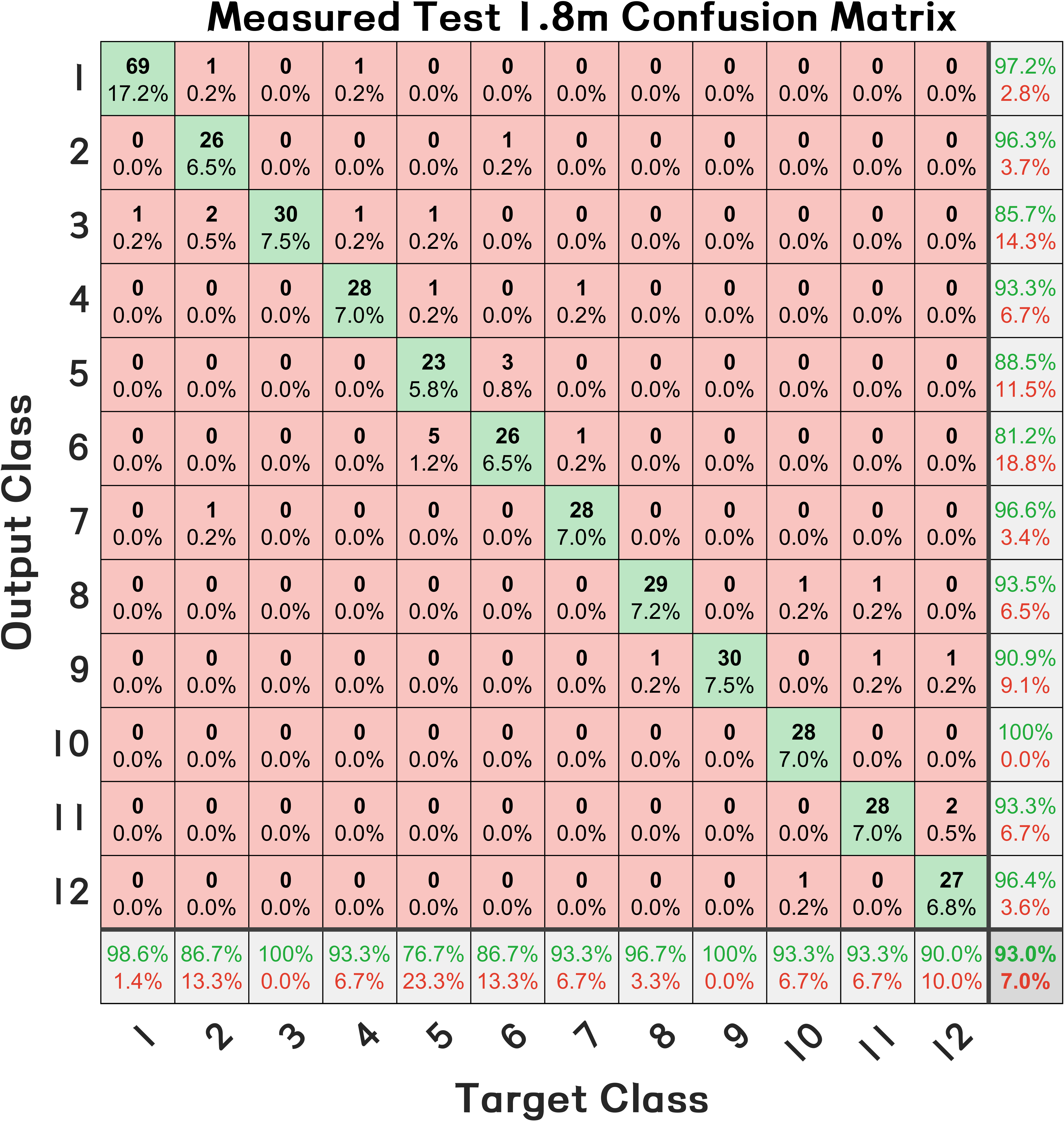}}
    \subfigure[]{\includegraphics[width=0.328\textwidth]{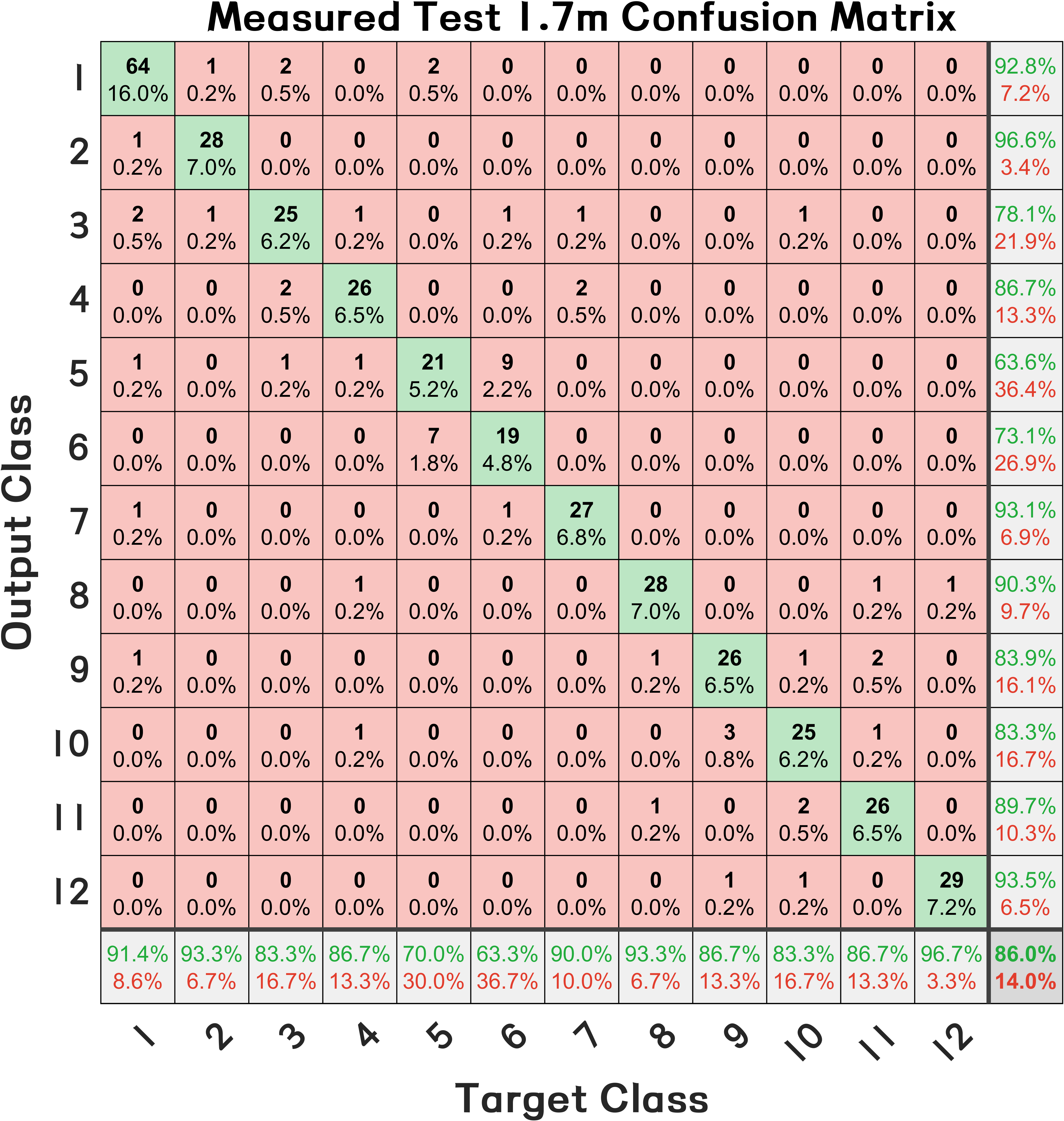}}
    \subfigure[]{\includegraphics[width=0.328\textwidth]{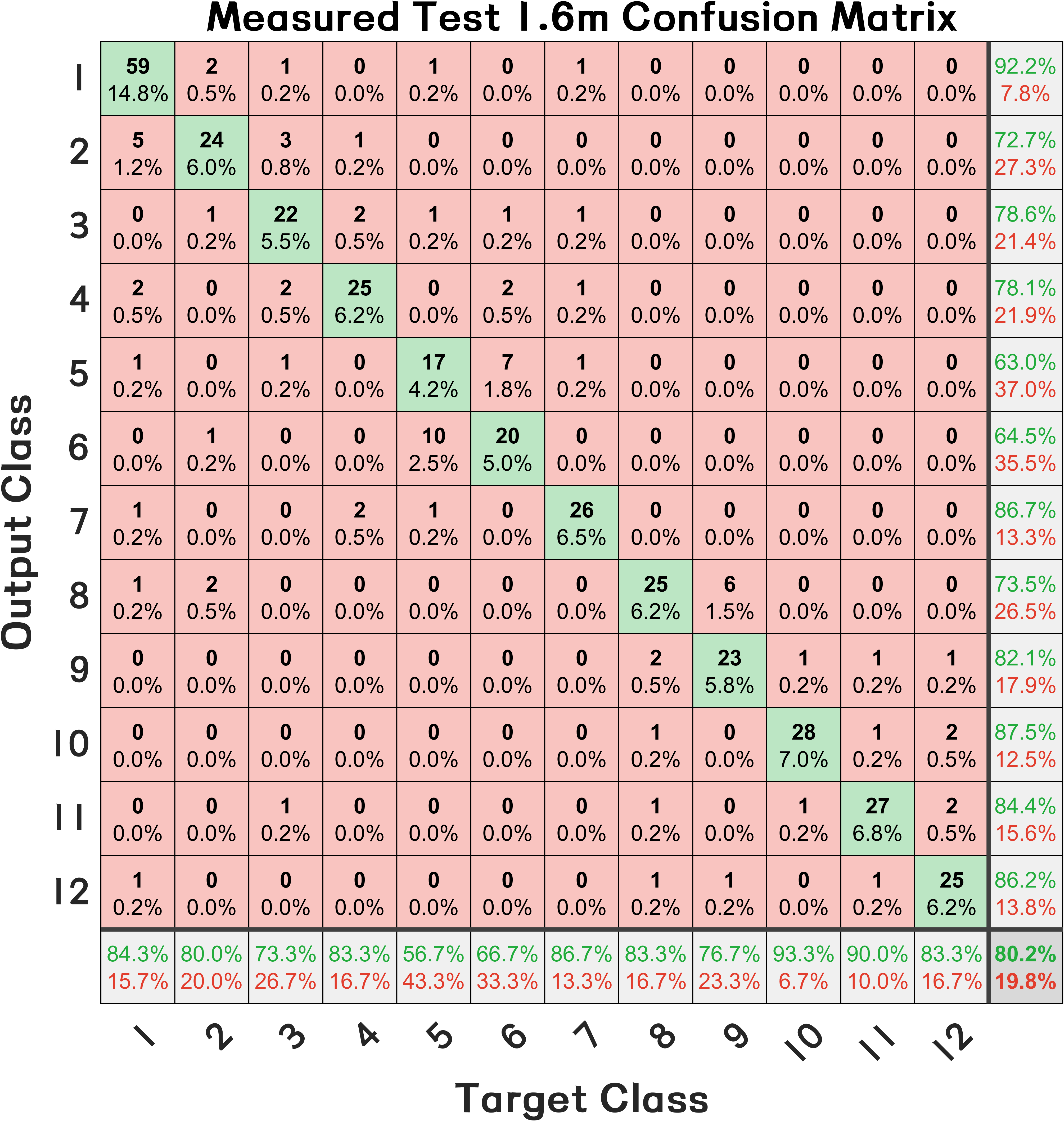}}
    \caption{Confusion matrices of the proposed method on validation and test sets: (a) Simulated validation, (b) Simulated test for $1.8~m$ human, (c) Simulated test for $1.7~m$ human, (d) Simulated test for $1.6~m$ human, (e) Simulated test for $1.5~m$ human, (f) Measured validation, (g) Measured test for $1.8~m$ human, (h) Measured test for $1.7~m$ human, (i) Measured test for $1.6~m$ human. The vertical axis in each matrix represents the predictions, the horizontal axis represents the target labels, and numbers $1\sim 12$ corresponds to the labels $S1\sim S12$.}
    \label{Confusion Matrices}
    \vspace{-0.0cm}
\end{figure*}\par
\subsection{Performance of the Proposed Method}
First, the $\mathbf{R^2TM}$, $\mathbf{D^2TM}$, polynomial-fit smoothed images, and 3D point cloud images are visualized and presented in Fig. \ref{Simulated Visualization} and \ref{Measured Visualization}. Fig. \ref{Simulated Visualization} shows the results on simulated data set. Each row contains $12$ images corresponding to the categories $S1 \sim S12$. By comparing the first row and the third row, it can be concluded that the extracted corners fitted features can represent the raw micro-Doppler information, while the noise is suppressed to a certain degree. The similar conclusion can be drawn by comparing the second and the forth row. By observing the $\mathbf{PC-RD}$ 3D corner point cloud after feature filtering and fusion, it is evident that the maps correspond to different categories have large feature separation. Fig. \ref{Measured Visualization} shows the results on measured data set. The results also prove that the 3D corner point cloud maps are helpful for recognition.\par
\begin{table*}[!ht]
\begin{center}
\caption{Comparison of the Accuracy of the Proposed Method with Existing Models$^{*}$.\label{Comparison Acc}}
\vspace{-0.0cm}
\resizebox{\textwidth}{!}{
\begin{tabular}{cccccccccccccc}
\hline\hline
\multirow{2}{*}{\textbf{Name of Methods}} & \multicolumn{7}{c}{\textbf{Simulated Data}} & \multicolumn{6}{c}{\textbf{Measured Data}} \\
\cline{2-14} & \textbf{Tr} & \textbf{Va} & \textbf{Va F-1} &\textbf{Te $1.8~m$} & \textbf{Te $1.7~m$} & \textbf{Te $1.6~m$} & \textbf{Te $1.5~m$} & \textbf{Tr} & \textbf{Va} & \textbf{Va F-1} & \textbf{Te $1.8~m$} & \textbf{Te $1.7~m$} & \textbf{Te $1.6~m$}\\
\hline
\multicolumn{14}{c}{\textbf{Frontier Methods in Computer Vision}}\\
\hline
ViT \cite{ViT} &$93.53$	&$89.88$	&$0.91$	&$89.75$	&$77.50$	&$\backslash$	&$\backslash$	&$94.69$	&$89.63$	&$0.90$	&$89.25$ & $77.75$ & $\backslash$
\\            
ConvNeXt \cite{ConvNeXt} &$94.06$	&$88.63$	&$0.92$	&$88.50$	&$78.75$	&$\backslash$	&$\backslash$	&$95.88$	&$89.25$	&$0.90$	&$87.75$ & $73.00$ & $\backslash$
\\          
\hline
\multicolumn{14}{c}{\textbf{Methods in Current Field without Feature Dimension Reduction}}\\
\hline
TWR-AEN-BiGRU$^{1}$ \cite{TWR-AEN-BiGRU} &$92.44$	&$82.50$	&$0.83$	&$80.25$	&$\backslash$	&$\backslash$	&$\backslash$	&$86.38$	&$80.75$	&$0.80$	&$81.50$ &$\backslash$ &$\backslash$
\\             
TWR-GCN$^{2}$ \cite{Wang 3} &$92.84$	&$85.75$	&$0.88$	&$85.25$	&$75.50$	&$\backslash$	&$\backslash$	&$92.03$	&$86.88$	&$0.88$	&$86.75$ & $79.50$ & $72.50$
\\  
TWR-ResNeXt$^{3}$ \cite{Chen 2} &$97.63$	&$91.25$	&$0.92$	&$92.00$	&$84.00$	&$72.75$	&$\backslash$	&$94.66$	&$89.63$	&$0.89$	&$88.50$ & $78.25$ & $\backslash$
\\   
TWR-CapsuleNet$^{4}$ \cite{Wang 2} &$100.00$	&$96.63$	&$0.95$	&$95.50$	&$86.25$	&$74.00$	&$\backslash$	&$99.19$	&$94.13$	&$0.94$	&$93.25$ & $82.25$ & $70.50$ 
\\  
\hline
\multicolumn{14}{c}{\textbf{Methods in Current Field with Feature Dimension Reduction}}\\
\hline
$\mu$D-PointNet$^{5}$ \cite{mD-PointNet} &$95.88$	&$90.25$	&$0.91$	&$90.75$	&$82.75$	&$77.50$	&$71.25$	&$96.47$	&$90.50$	&$0.91$	&$90.50$ & $82.75$ & $74.50$
\\            
RPCA-ResNet$^{6}$ \cite{An} &$88.41$	&$84.00$	&$0.85$	&$85.25$	&$80.00$	&$76.25$	&$69.50$	&$87.72$	&$81.25$	&$0.83$	&$88.25$ & $80.25$ & $74.00$
\\            
TWR-WSN-CRF \cite{TWR-WSN-CRF} &$99.50$	&$95.50$	&$0.95$	&$95.25$	&$84.25$	&$76.50$	&$\backslash$	&$99.34$	&$94.50$	&$0.94$	&$94.25$ & $80.50$ & $70.00$
\\   
\textbf{Proposed Method} &$\mathbf{99.09}$	&$\mathbf{94.63}$	&$\mathbf{0.94}$	&$\mathbf{94.50}$	&$\mathbf{90.00}$	&$\mathbf{87.00}$	&$\mathbf{78.50}$	&$\mathbf{98.44}$	&$\mathbf{93.38}$	&$\mathbf{0.94}$	&$\mathbf{93.00}$  &$\mathbf{86.00}$  &$\mathbf{80.25}$
\\   
\hline\hline
\end{tabular}
}
\end{center}
\footnotesize $^{*}$ “Tr”, “Va", and “Te” are the abbreviations of “training accuracy”, “validation accuracy”, and “test accuracy”, respectively. “F-1” is the abbreviation of “F-1 Score”. All accuracy rates are in $\%$ unit and F-1 scores are in $1$ unit. The boxes in the table marked with a diagonal line indicate that the models do not converge under the current conditions of testing per $10$ batches of training.\\
\footnotesize $^{1}$ The first step of the method is image expanding into a vector, here we keep the input image scale uniformly $256\times 256$.To ensure convergence within the number of training epochs of the network, the length of the long-short term memory (LSTM) network is set to $128$.\\
\footnotesize $^{2}$ Referring to the method in the original paper, the model inputs are set to $3$ links.\\
\footnotesize $^{3}$ The original method considers transfer learning and multi-link model integration. Due to the limitations of our single-perspective TWR system, we only adopt the network design of a single link.\\
\footnotesize $^{4}$ After obtaining confirmation from the authors, for the sake of rigor, we use the complex form of the RTM matrix as the network input.\\
\footnotesize $^{5}$ The model and the method we proposed share a similar idea, but its feature dimension reduction module utilizes constant false alarm rate (CFAR).\\
\footnotesize $^{6}$ The network model chosen for recognizing the dimension-reduced feature maps in this method is ResNet-50.\\
\vspace{-0.0cm}
\end{table*}\par
\begin{figure*}[!ht]
    \centering
    \includegraphics[width=\textwidth]{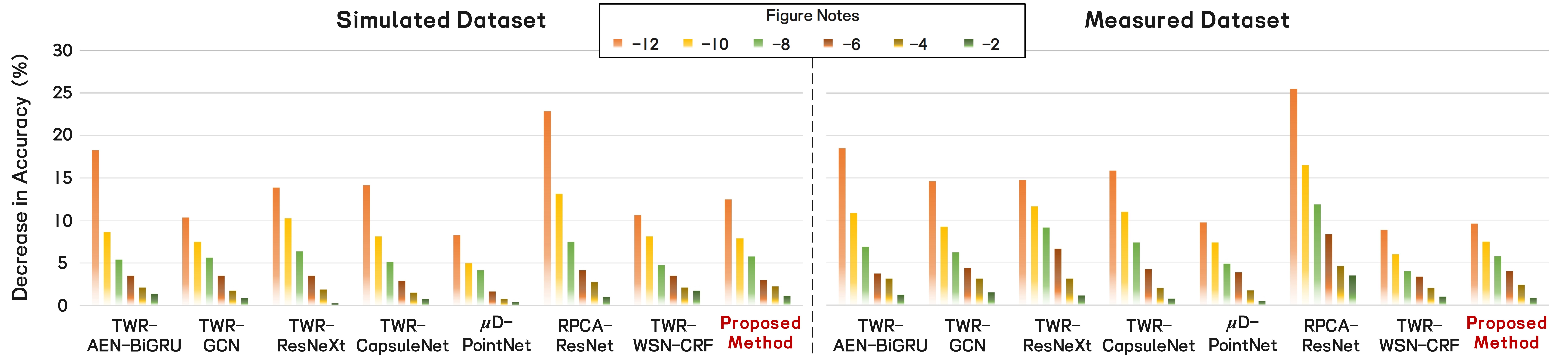}
    \caption{Robustness validation of the proposed method against existing methods, where the bar charts of different colors represent the magnitude of the decrease in SNR. The quantity is indicated in “Figure Note” in dB unit.}
    \label{Robustness Testing}
    \vspace{-0.0cm}
\end{figure*}\par
Secondly, confusion matrices are calculated on validation and test data sets to verify the effectiveness of the proposed method. The results are shown in Fig. \ref{Confusion Matrices}. In each confusion matrix, the vertical axis represents the predicted output labels and the horizontal axis represents the target labels. The green boxes on the diagonal represent the number of positive samples and the accuracy of the model for each category. The value in the lower right corner of the matrix is the total accuracy. The proposed method achieves a validation accuracy of $94.63\%$ on simulated data, and $93.38\%$ on measured data, implying that the model is effective and does not suffer from significant overfitting problems. The test accuracies on simulated data sets are $94.5\%$, $90.0\%$, $87.0\%$, and $78.5\%$ corresponding to $1.8$, $1.7$, $1.6$, and $1.5$ meters. Similarly, the test accuracies on measured data sets are $93.0\%$, $86.0\%$, and $80.2\%$ corresponding to $1.8$, $1.7$, and $1.6$ meters. It can be drawn that the accuracy decreases with the difference of heights between testers increases, which is consistent with the assertions drawn from kinematic modeling. The accuracies on measured data sets are lower than those on simulated data sets. From the perspective of different activities, the recognition accuracy of the proposed method on the two types of activities, $S5$ and $S6$, is generally lower than that on the other activities. This is due to the fact that the two types of activities contain only micro-Doppler information without any obvious macro-Doppler signature \cite{TWR-FMSN}. However, the validation accuracy on $S5$ and $S6$ is still not less than $80\%$, both for simulated and measured data.\par
\begin{table}
\begin{center}
\caption{Ablation Validation of Data Processing Module$^{*}$.\label{Data Processing Ablation}}
\vspace{-0.0cm}
\resizebox{0.48\textwidth}{!}{
\begin{tabular}{ccccccc}
\hline\hline
\multirow{2}{*}{\textbf{Name of Maps}} & \multicolumn{3}{c}{\textbf{Simulated Data}} & \multicolumn{3}{c}{\textbf{Measured Data}} \\
\cline{2-7}& \textbf{Tr}    & \textbf{Va} & \textbf{Va F-1} & \textbf{Tr}    & \textbf{Va} & \textbf{Va F-1}\\ \hline
Only RTM$^{1}$     & $81.28$ & $82.50$ & $0.81$ & $83.00$ & $80.63$ & $0.81$ \\
Only DTM$^{1}$      & $87.22$  & $84.38$ & $0.85$ & $89.50$ & $85.13$ &  $0.84$\\
Only $\mathbf{R^2TM}^1$    & $83.41$  & $79.88$ & $0.81$ & $81.91$ & $82.00$ &  $0.82$  \\
Only $\mathbf{D^2TM}^1$    &  $90.19$   & $88.38$ & $0.88$ & $88.25$ & $87.75$ &  $0.88$  \\
RTM \textcircled{c} DTM$^{2}$    &   $93.97$    & $90.13$ & $0.89$ & $89.97$ & $87.88$ &  $0.88$ \\
$\textbf{Proposed}^{3}$ & $\mathbf{99.09}$  & $\mathbf{94.63}$ & $\mathbf{0.94}$ & $\mathbf{98.44}$ & $\mathbf{93.38}$ & $\mathbf{0.94}$            \\
\hline\hline
\end{tabular}
}
\end{center}
\footnotesize $^{*}$  The definitions of all abbreviations are consistent with TABLE \ref{Comparison Acc}. All accuracy rates are in $\%$ unit and F-1 scores are in $1$ unit.\\
\footnotesize $^{1}$  The data input to the recognition network is still three-dimensional, but the default dimension is replaced by the number $0$.\\
\footnotesize $^{2}$  \textcircled{c} means fusing two images. The fusion method is consistent with the method proposed in the theoretical section.\\
\footnotesize $^{3}$  The proposed method is a fusion of both $\mathbf{R^2TM}$ and $\mathbf{D^2TM}$.
\vspace{-0.0cm}
\end{table}\par
\subsection{Comparison Experiments}
To evaluate the accuracy and generalization ability of the proposed method on TWR human activity recognition, $9$ existed models are employed for comparison, including: (1) Network methods used for image recognition at the frontiers of computer vision, including ViT and ConvNeXt; (2) Network methods used for TWR human activity recognition without artificial feature selection and dimension reduction procedure, including TWR-AEN-BiGRU, TWR-GCN, TWR-ResNeXt, and TWR-CapsuleNet; (3) Network methods used for TWR human activity recognition with artificial feature selection and dimension reduction procedure, including $\mu$D-PointNet, RPCA-ResNet and TWR-WSN-CRF.\par
As shown in TABLE \ref{Comparison Acc}, the proposed method achieves a validation accuracy of $94.63\%$ on the simulated data set and $93.75\%$ on the measured data set, which is close to the best existing model. With the decreasing of the testers' heights, the recognition accuracy decreases. Even worse some methods cannot work effectively. By comparison, almostly, the proposed method achieves the highest accuracy on test data sets, for example, $78.50\%$ on the $1.5~m$ tester in the simulation and $80.25\%$ on the $1.6~m$ tester in the experiment. The difference between the validation accuracy and the test accuracy reflects the generalization ability of the model. The difference is smaller, the generalization ability is better. The accuracy difference between the test set and the validation set is not more than $17\%$ in the simulation and $14\%$ in the experiment, which is the smallest among the existing methods. In addition, among all existing methods, although TWR-CapsuleNet has the highest accuracy on the validation set and the $1.8~m$ height test set, the proposed method still demonstrates better accuracy as the testers' height decreases.\par
\begin{table}
\begin{center}
\caption{Ablation Validation of Corner Detection and Filtering Module$^{*}$.\label{Corner Detection & Filtering Ablation}}
\vspace{-0.0cm}
\resizebox{0.38\textwidth}{!}{
\begin{tabular}{ccccc}
\hline\hline
\multirow{2}{*}{\textbf{Name of Maps}} & \multicolumn{2}{c}{\textbf{Simulated Data}} & \multicolumn{2}{c}{\textbf{Measured Data}} \\
\cline{2-5}& \textbf{Tr}    & \textbf{Va}  & \textbf{Tr}    & \textbf{Va} \\ \hline
Harris$^{1}$ \cite{Traditional Corner Detector}    & $90.44$ & $87.13$  & $82.94$ & $76.38$  \\
Moravec$^{1}$  \cite{Traditional Corner Detector}     & $87.81$ & $88.75$  & $85.91$ & $87.50$\\
FAST$^{1}$ \cite{Traditional Corner Detector}    & $94.25$ & $89.88$  & $90.53$ & $88.88$\\
DoG$^{2}$ \cite{DoG}   & $92.03$ & $86.50$  & $93.41$ & $90.00$\\
% DoG-$\mu$D-CornerDet w/o Filtering $^{2}$ \cite{Micro-Doppler Corner Detection}    & $95.59$ & $93.00$  & $93.88$ & $90.63$\\
$\textbf{Proposed}^{3}$ & $\mathbf{99.09}$  & $\mathbf{94.63}$  & $\mathbf{98.44}$ & $\mathbf{93.38}$ \\
\hline\hline
\end{tabular}
}
\end{center}
\footnotesize $^{*}$  The definitions of all abbreviations are consistent with TABLE \ref{Comparison Acc}. All accuracy rates are in $\%$ unit.\\
\footnotesize $^{1}$  These $3$ classes of methods are traditional image processing based corner detection algorithms. The number of filtered points is aligned with the kinematic model by a reasonable selection of thresholds. \\
\footnotesize $^{2}$  This method is a semi-finished design to the proposed method. The number of detected corners is obtained by direct control of the output through NMS and not through the filtering method.\\
\footnotesize $^{3}$  The proposed method contains the filtering method mentioned in the theoretical section.
\vspace{-0.0cm}
\end{table}\par
In addition, the noise robustness of the proposed method is verified. By adding Gaussian noise with different variances to the echo, it is possible to control the decrease of its image SNR from $-12\mathrm{~dB}$ to $-2\mathrm{~dB}$ in steps of $2\mathrm{~dB}$. All accuracy results are obtained on the aforementioned simulated and measured validation sets, respectively. In general, the decreasing trend of model validation accuracy with decreasing SNR and the number of training samples represents its corresponding robustness. The slower the accuracy decreases, the more robust the model is \cite{Radar Robustness}. As shown in Fig. \ref{Robustness Testing}, from the simulated set, the validation accuracy of the proposed method decreases by no more than $15~\%$ as the SNR decreases to $12\mathrm{~dB}$, which is close to the best existing methods. From the measured set, the validation accuracy of the proposed method decreases by no more than $10~\%$ as the SNR decreases to $12\mathrm{~dB}$, which is the best among all methods.\par
\subsection{Ablation Experiments}
In this subsection, three ablation experiments are conducted, including data processing methods, corner detection with filtering methods, and recognition methods.\par
First, whether or not stretching the vertical axes of the radar range and Doppler profiles into a square coordinate affects the final human activity recognition accuracy is verified. As shown in TABLE \ref{Data Processing Ablation}, the final training and validation accuracy obtained from the proposed data processing method is relatively larger than other comparison methods. Among the five data processing methods, the inputs for the first four are only one type of information from either the radar range or Doppler profiles. However, the input of the proposed network is a three-dimensional point cloud. In this case, the default dimension with all graph relations, weights and gradient information is meaningless, which greatly affects the validation accuracy of the recognition model. The results on the simulated data show that if only one type of radar profile input is used, the validation accuracy is about $1.7\%$ lower compared with the method of fusing two types of radar profiles. Similarly, using a single class of radar profile as input on measured data, the validation accuracy decreases by about $2.7\%$. With the exception of the radar range profile in the simulated data, stretching to a squared coordinate axis generally improves validation accuracy, and the positive impact on verification accuracy is greater after fusion. This is also in consistent with the human kinematics that squared coordinates are better to characterize the micro-Doppler information.\par
\begin{table}
\begin{center}
\caption{Ablation Validation of DGNN Module$^{*}$.\label{DGNN Ablation}}
\vspace{-0.0cm}
\resizebox{0.48\textwidth}{!}{
\begin{tabular}{ccccc}
\hline\hline
\multirow{2}{*}{\textbf{Name of Maps}} & \multicolumn{2}{c}{\textbf{Simulated Data}} & \multicolumn{2}{c}{\textbf{Measured Data}} \\
\cline{2-5}& \textbf{Tr}    & \textbf{Va}  & \textbf{Tr}    & \textbf{Va} \\ 
\hline
\multicolumn{5}{c}{\textbf{Graph Conducting Module}}\\
\hline
None   & $88.69$ & $83.63$  & $87.44$ & $81.13$  \\
Classic Graph Conductor$^{1}$  \cite{Traditional GNN}     & $96.63$ & $92.38$  & $93.94$ & $90.75$\\
$\textbf{Proposed}^{3}$ & $\mathbf{99.09}$  & $\mathbf{94.63}$  & $\mathbf{98.44}$ & $\mathbf{93.38}$ \\
\hline
\multicolumn{5}{c}{\textbf{Graph Focusing Module}}\\
\hline
Classic Graph Conv$^{2}$  \cite{Traditional GNN}     & $95.88$ & $90.13$  & $92.81$ & $88.63$\\
$\textbf{Proposed}^{3}$ & $\mathbf{99.09}$  & $\mathbf{94.63}$  & $\mathbf{98.44}$ & $\mathbf{93.38}$ \\
\hline
\multicolumn{5}{c}{\textbf{Graph Decision Module}}\\
\hline
MLP$^{3}$  \cite{Traditional Decision}     & $91.44$ & $88.88$  & $89.28$ & $82.38$\\
LSTM + Fc$^{4}$  \cite{Traditional Decision}     & $94.31$ & $91.50$  & $95.50$ & $89.75$\\
$\textbf{Proposed}^{3}$ & $\mathbf{99.09}$  & $\mathbf{94.63}$  & $\mathbf{98.44}$ & $\mathbf{93.38}$ \\
\hline\hline
\end{tabular}
}
\end{center}
\footnotesize $^{*}$  The definitions of all abbreviations are consistent with TABLE \ref{Comparison Acc}. All accuracy rates are in $\%$ unit.\\
\footnotesize $^{1}$  Simply achieves data-to-graph mapping using fully directed graphs.\\
\footnotesize $^{2}$  The classical layer design scheme in graph convolutional networks implements feature extraction through a combination of neighbor node aggregation, feature propagation, and nonlinear transformation.\\
\footnotesize $^{3}$  The essence is to use MLP to achieve decision-making.\\
\footnotesize $^{4}$  The essence is to use LSTM followed by a single-layer fully connected network to achieve decision-making.\\
\vspace{-0.0cm}
\end{table}\par
Second, we verify the effectiveness of the proposed corner detection and filtering methods. $4$ conventional image processing-based corner detectors are selected for comparison. Although it does not require an iterative training process, migrating the traditional corner detectors directly to the two types of radar images for feature extraction is not highly interpretable. For a trained network model, its inference time is close to that of traditional detectors, thus we consider this to be still an effective improvement. As shown in TABLE \ref{Corner Detection & Filtering Ablation}, keeping the back-end recognition decision model unchanged, the validation accuracy is no more than $90\%$ for traditional corner detectors on both simulated and measured data sets. The proposed method, by introducing $\mu$D-CornerDet and poly-fitted filtering methods, further improves the validation accuracy to $94.63\%$ on simulated data set and $93.75\%$ on measured data set.\par
Last, we verify the effectiveness of the proposed DGNN-based recognition decision module. The results are shown in TABLE \ref{DGNN Ablation}. The proposed module concatenates three different sub-modules in order, including graph constructing, graph focusing, and graph decision making. The graph constructing module employs spatial transform and the classical graph conductor is used for comparison. From TABLE \ref{DGNN Ablation}, compared with not using this module, the validation accuracy improves about $11\%$ while compared with using classical graph conductor, the validation accuracy improves about $2\%$. The graph focusing module employs edge convolution and the classical graph convolution is used for comparison. TABLE \ref{DGNN Ablation} shows that the validation accuracy improves about $4\%$. The graph decision module employs squeeze and excitation network. The multi-layer perceptron (MLP) and long-short-term-memory with fully-connected layer network (LSTM + Fc) are used for comparison. From TABLE \ref{DGNN Ablation}, the validation accuracy improves about $3\%$.\par

%%%%%%%%%%%%%%%%%%%%%%%%%%%%%%%%%%%%%%%%%%%%%%%%%%%%%%%%%%%%%%%%%%%%%%%%%%%%%%%%%%
% Conclusion
%%%%%%%%%%%%%%%%%%%%%%%%%%%%%%%%%%%%%%%%%%%%%%%%%%%%%%%%%%%%%%%%%%%%%%%%%%%%%%%%%%
\section{CONCLUSION}
This work has proposed a generalizable indoor human activity recognition method based on micro-Doppler corner point cloud and dynamic graph learning, to address the problems that existing methods generalize poorly to data under different testers. First, DoG-$\mathbf{\mu}$D-CornerDet has been used to extract micro-Doppler corner feature on two different kinds of radar profiles. Next, to maximize the feature distance within the constraints of the kinematic model, a polynomial fitting smoothing-based micro-Doppler corner filtering method has been proposed. Then, 3D point cloud has been obtained by concatenating the corners which have been taken from the two different types of radar profiles. Last, a data-to-activity label mapping method based on DGNN has been proposed for activity recognition. Numerical simulated and measured experiments on visualization, comparison, and ablation have been conducted to demonstrate the efficacy of the proposed method. The results have proved the generalization potential of the proposed method on radar data under various testers.\par

%%%%%%%%%%%%%%%%%%%%%%%%%%%%%%%%%%%%%%%%%%%%%%%%%%%%%%%%%%%%%%%%%%%%%%%%%%%%%%%%%%
% Acknowledgment
%%%%%%%%%%%%%%%%%%%%%%%%%%%%%%%%%%%%%%%%%%%%%%%%%%%%%%%%%%%%%%%%%%%%%%%%%%%%%%%%%%
% \section*{Acknowledgment}
% For a selection of permitted open source materials related to this research, please visit the author's official website at \url{https://weichenggaoresearch.company.site/}.\par
% The authors would like to thank Mr. Junbo Gong and Mr. Tian Lan from the Chongqing Innovation Center of Beijing Institute of Technology, Jiancheng Liao and Jingbo Wang from the Special Radar Laboratory (Formerly: the New System Radar Laboratory) of Beijing Institute of Technology for their help in related experiments. In addition, the authors would also like to thank Mr. Zhengliang Zhu and his team for their outstanding work on "Open-source Dataset of Human Motion Status Using IR-UWB Through-Wall Radar", Dr. Pengyun Chen and his team for their series of outstanding work on "Ultra-wideband radar human motion intelligent recognition", which are the inspiration and encouragement of this work.\par

%%%%%%%%%%%%%%%%%%%%%%%%%%%%%%%%%%%%%%%%%%%%%%%%%%%%%%%%%%%%%%%%%%%%%%%%%%%%%%%%%%
% References
%%%%%%%%%%%%%%%%%%%%%%%%%%%%%%%%%%%%%%%%%%%%%%%%%%%%%%%%%%%%%%%%%%%%%%%%%%%%%%%%%%

%%%%%%%%%%%%%%%%%%%%%%%%%%%%%%%%%%%%%%%%%%%%%%%%%%%%%%%%%%%%%%%%%%%%%%%%%%%%%%%
% Introductions of Authors
%%%%%%%%%%%%%%%%%%%%%%%%%%%%%%%%%%%%%%%%%%%%%%%%%%%%%%%%%%%%%%%%%%%%%%%%%%%%%%%
\newpage
\begin{IEEEbiography}[{\includegraphics[width=1in,height=1.25in,clip,keepaspectratio]{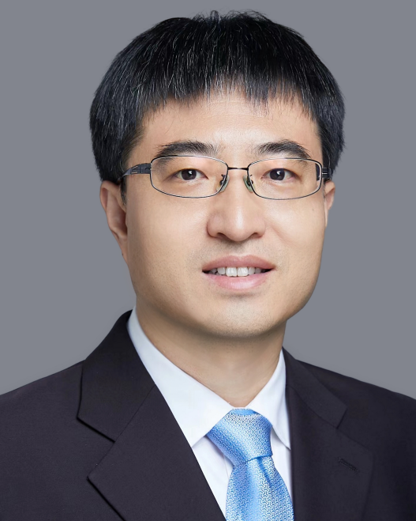}}]{Xiaopeng Yang}
Xiaopeng Yang (Senior Member, IEEE), professor, received B.E. and M.E. degrees from Xidian University, China, in 1999 and 2002, respectively, and Ph.D. degree from Tohoku University, Japan, in 2007. He was a Post-Doctoral Research Fellow with Tohoku University from 2007 to 2008 and a Research Associate with Syracuse University, USA, from 2008 to 2010. Since 2010, he has been working with the School of Information and Electronics, Beijing Institute of Technology (BIT), Beijing, China, where he is currently a Full Professor.\par
His current research interests include the phase array radar, ground-penetrating radar and through-the-wall radar.\par
\end{IEEEbiography}
\begin{IEEEbiography}[{\includegraphics[width=1in,height=1.25in,clip,keepaspectratio]{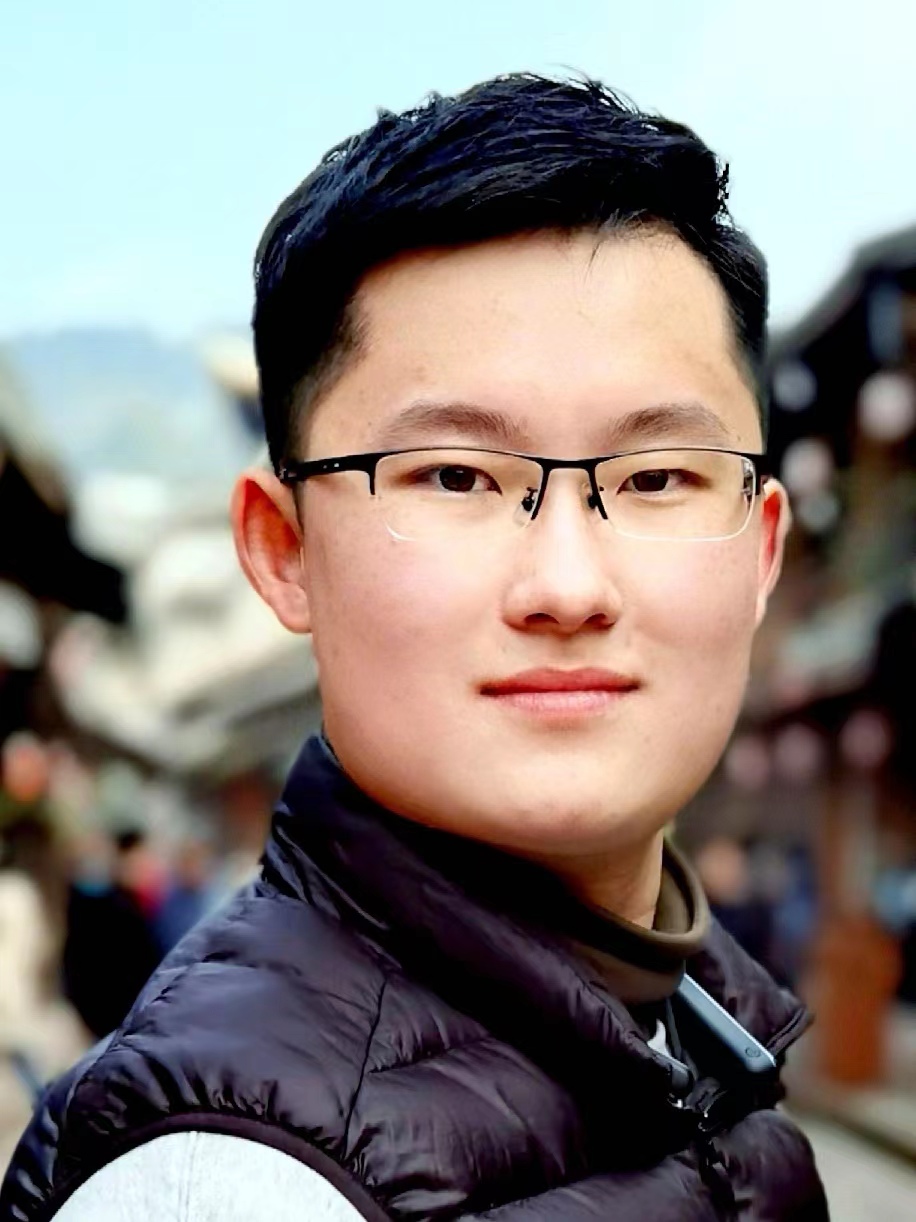}}]{Weicheng Gao}
Weicheng Gao (Graduate Student Member, IEEE), received his B.S. degree in Beijing Institute of Technology (BIT) in 2022. He is pursuing the Ph.D. degree at the Research Lab of Radar Technology, BIT. He is selected as a member of the China Association for Science and Technology (CAST) Talent Program.\par
His research interests are mainly focused on machine intelligence based through-the-wall radar (TWR) human micro-Doppler signature extraction, activity and gait recognition method.\par
\end{IEEEbiography}
\begin{IEEEbiography}[{\includegraphics[width=1in,height=1.25in,clip,keepaspectratio]{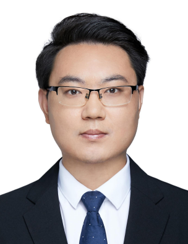}}]{Xiaodong Qu}
Xiaodong Qu (Member, IEEE), associate researcher, received the B.S. degree from Xidian University, Xi’an, China, in 2012, and the Ph.D. degree from the University of Chinese Academy of Sciences, Beijing, China, in 2017.\par
His research interests mainly include array signal processing, through-the-wall radar imaging.\par
\end{IEEEbiography}
\begin{IEEEbiography}
[{\includegraphics[width=1in,height=1.25in,clip,keepaspectratio]{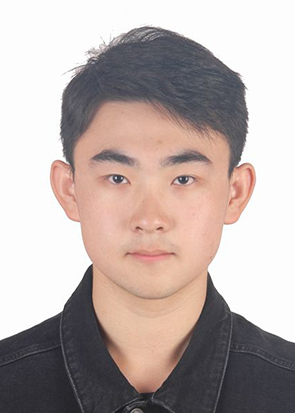}}]
{Haoyu Meng}
Haoyu Meng (Graduate Student Member, IEEE), was born in Taiyuan City, Shanxi Province, in 1999. He received the B.S. degree in Beijing Institute of Technology, Beijing, China, in 2021. He is currently pursuing his Ph.D. degree at Beijing Institute of Technology, Beijing, China. Since 2021. He is a student in the School of Information and Electronics, Beijing Institute of Technology.\par
His research interests include adaptive beamforming, mainlober interference suppression and through-the-wall radar.\par
\end{IEEEbiography}

%%%%%%%%%%%%%%%%%%%%%%%%%%%%%%%%%%%%%%%%%%%%%%%%%%%%%%%%%%%%%%%%%%%%%%%%%%%%%%%%%%
% The End of the Paper and Some Notes
%%%%%%%%%%%%%%%%%%%%%%%%%%%%%%%%%%%%%%%%%%%%%%%%%%%%%%%%%%%%%%%%%%%%%%%%%%%%%%%%%%
% \end{spacing}
\end{document}